\documentclass[twocolumn]{aastex631}
\pdfoutput=1 
\usepackage{amsmath,amssymb,amstext}
\usepackage{gensymb}
\usepackage{url}
\usepackage{comment}
\usepackage{tabularx}
\usepackage{booktabs}
\usepackage{array}
\usepackage[normalem]{ulem}
\usepackage{enumitem}

\def\lsim{\;\rlap{\lower 2.5pt
   \hbox{$\sim$}}\raise 1.5pt\hbox{$<$}\;}
\def\gsim{\;\rlap{\lower 2.5pt
 \hbox{$\sim$}}\raise 1.5pt\hbox{$>$}\;}

\usepackage{appendix}

\submitjournal{ApJ}

\shorttitle{Cosmic Ray Effects on Turbulence}
\shortauthors{Bustard \& Oh}

\begin{document}

\title{Cosmic Ray Drag and Damping of Compressive Turbulence}

\correspondingauthor{Chad Bustard}
\email{bustard@ucsb.edu}
\author[0000-0002-8366-2143]{Chad Bustard}
\affil{Kavli Institute for Theoretical Physics, University of California - Santa Barbara, Kohn Hall, Santa Barbara, CA 93107, USA}

\author{S. Peng Oh}
\affil{Department of Physics, University of California - Santa Barbara, Broida Hall, Santa Barbara, CA 93106, USA}

\begin{abstract}

While it is well-known that cosmic rays (CRs) can gain energy from turbulence via second order Fermi acceleration, how this energy transfer affects the turbulent cascade remains largely unexplored. Here, we show that damping and steepening of the compressive turbulent power spectrum are expected once the damping time $t_{\rm damp} \sim \rho v^{2}/\dot{E}_{\rm CR} \propto E_{\rm CR}^{-1}$ becomes comparable to the turbulent cascade time. Magnetohydrodynamic (MHD) simulations of stirred compressive turbulence in a gas-CR fluid with diffusive CR transport show clear imprints of CR-induced damping, saturating at $\dot{E}_{\rm CR} \sim \tilde{\epsilon}$, where $\tilde{\epsilon}$ is the turbulent energy input rate. In that case, almost all the energy in large scale motions is absorbed by CRs and does not cascade down to grid scale. Through a Hodge-Helmholtz decomposition, we confirm that purely compressive forcing can generate significant solenoidal motions, and we find preferential CR damping of the compressive component in simulations with diffusion and streaming, rendering small-scale turbulence largely solenoidal, with implications for thermal instability and proposed resonant scattering of $E \gsim 300$ GeV CRs by fast modes. When CR transport is streaming dominated, CRs also damp large scale motions, with kinetic energy reduced by up to to an order of magnitude in realistic $E_{\rm CR} \sim E_{\rm g}$ scenarios, but turbulence (with a reduced amplitude) still cascades down to small scales with the same power spectrum. Such large scale damping implies that turbulent velocities obtained from the observed velocity dispersion may significantly underestimate turbulent forcing rates, i.e. $\tilde{\epsilon} \gg \rho v^{3}/L$. 

\end{abstract}

\section{Introduction}
Cosmic rays (CRs) and magnetized turbulence are both ubiquitous in the Universe, and their interplay has long been a fascinating topic of research. Fluctuations at the small-scale end of a turbulent cascade, on scales of order the CR gyroscale, are frequently invoked to scatter individual CRs, creating the high degree of observed CR isotropy and the long residence times of CRs in the Milky Way disk and its surrounding halo relative to the light crossing time \citep{AmatoReview2018, Tjus2020}. In such a scenario, dubbed the ``extrinsic turbulence" model \citep{ZweibelReview2017}, the resulting bulk CR transport is magnetic field-aligned diffusion, with an energy-dependent spatial diffusion coefficient $\kappa_{||}$ and CR flux $F_{CR} \propto \kappa_{||} \nabla P_{CR}$. CRs in this picture can also gain energy from repeated scattering off gyroscale fluctuations, a second order Fermi mechanism called ``resonant reacceleration."

Phenomenological models of CR propagation fit to direct and indirect CR observables \citep{Hanasz2021} have traditionally assumed a Kolmogorov scaling for turbulence, appropriate for hydrodynamic turbulence; however, our understanding of CR scattering by turbulence has been refined over time with new insights into magnetohydrodynamic (MHD) turbulence. Most profoundly, MHD turbulence differs from hydrodynamic turbulence in that MHD forces and hence turbulence are no longer isotropic. 
The resulting anisotropy of slow and Alfv\'{e}n modes \citep{GoldreichSridhar1995} makes them inefficient CR scatterers, as CRs interact with multiple uncorrelated eddies during one gyro-orbit, essentially canceling out gyroresonant contributions from each eddy \citep{Chandran2000}. 

Compressible fast modes, whose velocities are independent of magnetic field direction, are more isotropic \citep{cho2003} and therefore considered the best candidate for CR scattering \citep{Yan2004}; although, the degree of isotropy decreases with decreasing scale due to strong collisionless and viscous damping, hence the efficacy of CR scattering decreases with decreasing CR energy \citep{Kempski2022}. Fast mode scattering, then, is most plausible for higher energy CRs ($E > 300$ GeV). 

For $E < 300$ GeV, where most of the CR energy resides, CRs can largely create scattering perturbations themselves through a resonant streaming instability \citep{Wentzel1968, Kulsrud1969}. The resulting transport is no longer purely diffusive; instead, CRs ``stream" down their field-aligned pressure gradient at the local Alfv\'{e}n speed $v_{A} = B/\sqrt{4\pi\rho}$ with $F_{CR} \propto v_{A} P_{CR}$, and additional, energy-dependent CR diffusivity ($F_{CR} \propto \nabla P_{CR}$) is introduced by wave damping\footnote{Note that, depending on the functional form of the damping rate, the ``diffusive" term may not be truly diffusive (see e.g. \citealt{Skilling1971} or Appendix B3 of \citealt{Hopkins2021_testing} for examples)}, e.g. ion-neutral damping, nonlinear Landau damping, and turbulent damping \citep{Skilling1971, Farmer2004, Blasi2012, WienerWIM, ZweibelReview2017, Bustard2021}. There is also an important difference regarding energy transfer between CRs and hydromagnetic waves: whereas extrinsic turbulence is generated externally, in self-confinement, the free energy to generate waves comes from the CRs themselves, and this energy is subsequently dissipated into the thermal gas via wave damping at a rate $H = -dE_{CR}/dt = v_{A} \cdot \nabla P_{CR}$. We refer to this collisionless energy transfer as streaming energy loss / gas heating. 


While considerable effort has been put towards exploring resonant-scale interactions between CRs and either self-generated (e.g. \citealt{Skilling1975, Felice2001, bai2019, holcomb2019}) or externally driven (e.g. \citealt{giacalone1999, Yan2002, Reichherzer2020}) waves, somewhat less focus has been given to the interplay between CRs and turbulence on scales much larger than a CR gyroradius (less than an AU for a GeV CR proton in a $\sim \mu$G field). In particular, we will focus on scales larger than the CR mean free path due to pitch angle scattering\footnote{This is usually around a pc both in phenomenological models of Milky Way CR propagation motivated by extrinsic turbulence and in self-confinement models.}, where the collective CR population is well-described as a fluid that experiences compressions and rarefactions in the turbulent flow, leading to energy transfer between the CRs and turbulence. To distinguish this from its resonant-scale counterpart, the flow of energy from turbulence to the bulk CR fluid is called non-resonant reacceleration \citep{Ptuskin1988}, and its efficiency depends on CR transport model. 

For purely diffusive CR transport, 
non-resonant reacceleration is maximally efficient when CRs are well-trapped in the turbulent flow ($\kappa < v_{ph} L_{0}$, where $v_{ph}$ is the phase speed of compressive fluctuations and $L_0$ is the outer eddy scale). When streaming is taken into account, the interaction between perturbed CR and gas variables is fundamentally altered. While CR diffusion introduces a $\pi/2$ phase shift between CR and density perturbations, leading to a CR force that damps fluctuations much like a damped harmonic oscillator, both the change in flux ($F_{CR} \propto P_{CR}$ instead of $F_{CR} \propto \nabla P_{CR}$) and the associated energy loss that accompany streaming transport modify the CR force \citep{Tsung2021_staircase}. As we showed in \cite{BustardOh2022_reacceleration} (from now on referred to as Paper I), CR reacceleration / turbulent damping rates become dependent on plasma $\beta = P_{g}/P_{B}$; they remain largely unchanged in high-$\beta$ plasmas like the intracluster medium (ICM) where reacceleration is a leading explanation for radio halos (e.g. \citealt{Brunetti2011, BrunettiJones2014_review}), but they are stunted significantly in low-$\beta$ plasmas. 

Despite non-resonant reacceleration being a fairly inefficient process compared to diffusive shock acceleration (a first order Fermi mechanism), with minimum growth times lengthened even further by streaming transport, it was pointed out by \cite{Thornbury2014, Drury2017} that a significant fraction of total CR power in galaxies could come from reacceleration, consequently creating a large sink for turbulent energy. In this paper, we present analytical estimates and CR+MHD simulations suggesting that CRs in very plausible astrophysical environments can divert significant amounts of turbulent energy, essentially acting as an unsual form of viscosity. The outcome is a CR-modified route to gas heating, rather than the typical conversion to heat at the dissipation scale, and a damped turbulent energy spectrum with decreased small-scale, compressive power. 

These changes are, of course, strongest in environments where CRs are dynamically important such as the ISM (where CR energy densities are roughly in equipartition with turbulent and magnetic energy densities; \citealt{Boulares1990}) and the Milky Way circumgalactic medium (which may be energetically dominated by CRs; e.g. \citealt{JiCRHalos2020}), but they would affect any process that relies on compressive motions. For instance, compressions seed thermal instability \citep{Field1965, McCourt2012, Mohapatra2022}, which is frequently invoked, for instance, to explain the existence of cold CGM clouds \citep{Putman2012_review}. Fluctuations that scatter CRs are not immune to these modifications either. Low-energy, self-confined CRs could sap energy from the turbulent fast mode cascade at large scales, decreasing the available small-scale power needed to scatter \emph{high} energy CRs.

This paper is outlined as follows. In \S \ref{sec:setup}, we discuss our simulation method and setup. In \S \ref{sect:CR-diversion}, we analytically estimate and then quantify in simulations the fractions of turbulent driving and gas heating that are channeled through CRs. We then analytically derive how CR-induced damping should affect MHD turbulence spectra (\S \ref{sec:damping_analytics}) and the conditions under which damping rates can exceed cascade rates (\S \ref{sec:tcascade}). In \S \ref{sec:damping-sims}, we present exploratory simulations strongly suggestive of these analytic estimates and show sensitivities to streaming vs diffusive CR transport. We discuss regimes of applicability and implications in \S \ref{sec:discussion} and conclude in \S \ref{sec:conclusions}.

\section{Simulation Setup}
\label{sec:setup}

\begin{table*}
  \centering
  \caption{Simulation parameters, CR module settings, and other variable definitions}
  \begin{tabular}{|l|l|r|}
    \toprule \toprule
    {\bf Parameter} & {\bf Definition / Setting / Equation} & {\bf Additional Notes} \\
    \midrule
    L & Half box size & k = 2 mode  \\
    
    $L_0$ & Outer eddy scale & k = 3 mode \\
    
    $t_{\rm drive}$ & $2 \times 10^{-3} (L/c_{s})$ & Turbulence driven every $t_{\rm drive}$   \\
    
    $t_{\rm corr}$ & $L/c_{s}$ & Autocorrelation time \\
    
    $\tilde{\epsilon}$, $\epsilon$  & Input turbulent energy rate, dE/dt & $\rho v^{3}/L$, $v^{3}/L$ in hydro turbulence \\
    
    $v_{m}$ & $50c_{s}$ & Effective maximum speed of light \\

    $\kappa$ & CR diffusion coefficient & Assumed to be field-aligned only ($\kappa = \kappa_{||}$) \\
    
    $\beta$ & $P_{\rm g}/P_{B}$ & Plasma beta \\
    
    $c_{s}$ & $\sqrt{\gamma P_{g}/\rho}$ & Gas sound speed \\
    $v_{ph}$ & $\sqrt{(\gamma P_{g} + \gamma_{CR} P_{\rm CR} + P_{B})/\rho}$ & Compressive wave phase speed \\ 
    $v_{A}$ & $B/\sqrt{4\pi \rho}$ & Alfv\'{e}n speed \\ 
    $c_{c}$ & $\sqrt{\gamma_{CR} P_{CR}/\rho}$ & Effective CR sound speed \\
    
    $\mathcal{M}_{\rm s}$, $\mathcal{M}_{\rm ph}$, $\mathcal{M}_{\rm A}$, $\mathcal{M}_{\rm c}$ & $v/c_{s}$, $v/v_{ph}$, $v/v_{A}$, $v/c_{c}$ & Mach numbers \\

    H & $v_{A} \cdot \nabla P_{CR}$ & ``Collisionless" CR loss rate / gas heating rate \\
    
    $f_{\rm CR}$, $f_{\rm th}$, $f_{\rm CR, heating}$ & $\dot{E}_{CR}/\tilde{\epsilon}$, $\dot{E}_{th}/\tilde{\epsilon}$, $<H>$/$\tilde{\epsilon}$ & Fraction of $\tilde{\epsilon} \rightarrow$ CRs, thermal gas, CR heating   \\

    E(k) & Kinetic energy spectrum & $\propto k^{-5/3}$ (Kolmogorov), $k^{-2}$ (Burgers), $k^{-3/2}$ (Kraichnan) \\

    $t_{\rm inject}$ & $\rho v^{2}/\tilde{\epsilon}$ & Energy injection time \\
    
    $t_{\rm cascade}$ & $k E(k)/F(k)$ & Cascade time (see Equation \ref{eq:tcascade_k}) \\

    $t_{\rm grow}$ & $p^{2}/D_{pp}$ & CR reacceleration time (\S \ref{sect:CR-diversion} and Paper I) \\ 

    $t_{\rm damp}$ & $\sim \rho v^{2} \rm max \left( \frac{t_{\rm grow}}{P_{\rm CR}}, \frac{1}{\tilde{\epsilon}} \right) \sim \rm max \left(\mathcal{M}_{\rm c}^{2} t_{\rm grow},t_{\rm inject} \right)$ & Turbulent damping time (Equation \ref{eq:tdamp_gen}) \\

    \bottomrule
  \end{tabular}
\label{param_table}
\end{table*}

We begin by briefly describing the simulation methodology and setup, which is described in more detail in Paper I. Using the Athena++ MHD code \citep{AthenaRef} coupled with an additional CR module that models CR diffusive and streaming transport in a fluid approximation using a two-moment method originally developed for radiation transport \citep{JiangCRModule}, we numerically solve the ideal MHD equations plus two additional equations for the CR energy and energy flux. All simulations begin with a flat background (no gradients) consisting of CRs, gas, and magnetic fields, with a constant net (straight) magnetic field in the $\hat{x}$ direction. We stir turbulence following an Ornstein-Uhlenbeck random process \citep{Uhlenbeck1930, EswaranPope1988}, randomly generating velocity perturbations between modes k = 1 and 3 in a cubic box of width 2L. For driving, we set the autocorrelation timescale to be $t_{\rm corr} =  L/c_{s}$ and drive fluctuations every $t_{\rm drive} = 2 \times 10^{-3} (L/c_{s})$. For the parameter scans in \S \ref{sect:CR-diversion}, we use grids of size $128^{3}$ and $256^{3}$. We simulate fluids with either an isothermal equation of state, where the thermal energy is fixed, or an adiabatic equation of state. The latter results in a gradual rise in the gas pressure due to a combination of CR heating and grid-scale dissipation of the cascade, which we decompose and quantify. These simulations all use purely compressive forcing, with two turbulent driving rates $\tilde{\epsilon} = dE/dt$, resulting in approximately $\mathcal{M}_{s} \sim 0.15$ and $\mathcal{M}_{s} \sim 0.5$ turbulence with a weak dependence on plasma $\beta$ since MHD forces 
counteract motions. We avoid solenoidal driving to avoid turbulent amplification of magnetic fields, so that we can evolve simulations at approximately fixed plasma $\beta$. To a good approximation, solenoidal driving only amplifies magnetic fields, while compressive driving energizes CRs.

At our parameter scan resolution of 2L/256, the cascade exhibits only a short inertial range, and in testing we find that the spectral slope in pure MHD runs (no CRs) is intermediate between $E(k) \sim k^{-2}$ and $E(k) \sim k^{-3/2}$ -- a shallower slope is expected for compressive fast modes, but the exact exponent has been highly debated. In our analytic estimates (\S \ref{sec:damping_analytics}), we will explore CR-induced deviations to different initial spectra, but we particularly note significant changes to Kraichnan turbulence where $E(k) \sim k^{-3/2}$ initially. For \S \ref{sec:damping-sims}, where we want to test deviations from this spectrum due to CR drag, we increase the resolution to 2L/512, though we find that the main trends are well-recovered even with a resolution of 2L/256 (see Appendix). Higher resolution simulations giving a larger inertial range would be preferable, but to ensure an accurate treatment of CR propagation and influence, the two-moment method has an effective, maximum speed of light parameter $v_m$ that must be much larger than other propagation speeds in the system and that sets the Courant-limited timestep. In Paper I, we found that $v_m \sim 50 c_s$ gives seemingly converged CR heating rates and reacceleration rates. With this choice, our MHD+CR simulations are about a factor of 8 more expensive than pure hydro turbulence sims, prohibiting us from going to much higher resolution.

\section{Cosmic Ray Diversion of Turbulent Energy}
\label{sect:CR-diversion} 

We'll begin with a short review of non-resonant reacceleration (see e.g. \citealt{Ptuskin1988, Chandran2004, Lynn2012} and \S 2 of Paper I for greater detail) and its relation to the turbulent damping rate. Variables used in our discussion are summarized in Table \ref{param_table}. As discussed in Paper I, ``drag" against CRs provides a frictional force on compressive motions known as Ptuskin damping \citep{ptuskin81}. It is similar to radiative damping of sound waves, which famously leads to Silk damping of acoustic waves in the early universe \citep{Silk1968}. In general, since $E_{\rm k}/t_{\rm damp} \sim  P_{\rm CR}/t_{\rm grow}$, we have\footnote{In this paper, we use the notation $\tilde{\epsilon}$ to denote the turbulent driving rate in units of turbulent energy density per unit time, and we use $\epsilon$ to denote the driving rate in units of  
$v^2$ (velocity squared) per unit time. In hydrodynamic turbulence, $\tilde{\epsilon} \equiv \rho v^{3}/L$ and ${\epsilon} \equiv v^{3}/L$, but these equivalences don't hold in CR-modified turbulence.}:
\begin{equation} 
t_{\rm damp} \sim \rho v^{2} \rm max \left( \frac{t_{\rm grow}}{P_{\rm CR}}, \frac{1}{\tilde{\epsilon}} \right) \sim \rm max \left(\mathcal{M}_{\rm c}^{2} t_{\rm grow},t_{\rm inject} \right)
\label{eq:tdamp_gen} 
\end{equation}
where $\mathcal{M}_{\rm c} \equiv v/c_c$ is the Mach number in units of the CR effective sound speed, $c_c \sim \sqrt{P_{\rm CR}/\rho}$, and $t_{\rm inject} \equiv \rho v^{2}/\tilde{\epsilon}$. Equation \ref{eq:tdamp_gen} is a general expression for the damping time, for which one can plug in the appropriate $t_{\rm grow}$, the CR reacceleration (or growth) time.

Working in the limit of purely diffusive spatial CR transport with isotropic diffusion coefficient $\kappa$, the reacceleration time can be derived in two limits depending on the ratio of diffusion time $t_{\rm diff} = l^{2}/\kappa$ to compressive wave crossing time $t_{\rm sc} = l/v_{ph}$ across an eddy of length $l$ in a medium with compressive phase velocity $v_{\rm ph} \sim (P_{\rm tot}/\rho)^{1/2} \sim [P_{\rm g} + P_{\rm B} + P_{\rm CR})/\rho]^{1/2}$. In the fast diffusion limit ($t_{\rm diff} \ll t_{\rm sc}$, or equivalently, $\kappa \gg v_{ph}l$), deriving the CR momentum diffusion coefficient $D_{\rm pp}$ follows the textbook argument for second order Fermi acceleration: $D_{\rm pp} \sim (\Delta p)^{2} / \tau_{\rm scatter} \sim p^{2} v^{2}/(c^{2} \tau_{\rm scatter}) \sim p^{2} v^{2} / \kappa$. The energy growth time, defined as $p^{2}/D_{pp}$ is 
\begin{equation}
    t_{\rm grow} \sim \frac{\kappa}{v^{2}}; \quad \kappa >> v_{ph}l
    \label{eqn_fastDiffusion}
\end{equation}

In the opposite limit of slow diffusion ($t_{\rm diff} \gg t_{\rm sc}$, or equivalently, $\kappa \ll v_{ph}l$), $D_{pp} \sim (\delta p)^{2}/\tau_{\rm diff} \sim (p^{2} v^{2}/v_{\rm ph} ^2) (\kappa/l^{2})$, and the growth time is
\begin{equation}
    t_{\rm grow} \sim \frac{p^{2}}{D_{pp}} \sim \frac{v_{\rm ph}^{2} l^2}{v^{2} \kappa}; \quad \kappa << v_{\rm ph}l
    \label{eqn_slowDiff}
\end{equation}
Joining the two regimes in the middle, the minimum growth time is $t_{\rm grow} \sim (v_{\rm ph} l/v^2)$ when $\kappa \sim v_{ph} l$.

Strictly speaking, these scalings are appropriate if CR diffusion is isotropic, if streaming is negligible, and if all reacceleration comes from eddies of a single scale $l$. Relaxing these assumptions introduces further modifications. In the fast diffusion limit ($\kappa \gg v_{ph} l$), there are also correction factors that decrease the growth time if anisotropic rather than isotropic spatial diffusion is accounted for \citep{Chandran2004}. Additional streaming transport, widely applicable for CRs with energy $E \lessapprox 300$ GeV, introduces a correction factor that decreases reacceleration rates by $f_{\rm corr} = 1 - \sqrt{2/ \beta}$ and $f_{\rm corr} = (1 - \sqrt{2/ \beta})^{1/2}$ in the slow and fast diffusion regimes, respectively (Paper I); and in the slow diffusion limit ($\kappa \ll v_{ph} l$), multiple eddies contribute to reacceleration, with relative contributions dependent upon the shape of the turbulent power spectrum (see Equation 4 in Paper I for a more general expression). If the wave spectrum is Burgers-like ($E(k) \sim k^{-2}$), roughly consistent with our simulations, eddies at each logarithmic interval in the inertial range contribute equally to reacceleration, and $t_{\rm grow}$ has a broad minimum of $t_{\rm grow} \sim (v_{\rm ph} l/v^2)$ throughout the entire range of $\kappa_{||} < v_{ph} l$.  

If we work in the limit of a single outer-scale eddy (i.e., we only consider eddies of size $L_0)$, in the fast diffusion ($\kappa \gg v_{\rm ph} L_0$) regime, where $t_{\rm grow} \sim \kappa/v^{2}$ then Equation \ref{eq:tdamp_gen} gives $t_{\rm damp} \sim \kappa/c_c^2$, in agreement with the classic (much more detailed) calculation of this effect by \citet{ptuskin81}. Working instead in the broad regime of maximal reacceleration, where CRs are well-trapped in the turbulent flow (when $\kappa < v_{ph} L_{0}$), the characteristic growth time is $t_{\rm grow} \sim (v_{\rm ph} L_0/v^2)$, which gives: 
\begin{equation}
    t_{\rm damp} \sim \rm max  \left( \frac{v_{\rm ph} L_0}{c_c^2}, t_{\rm inject} \right)
\label{eq:tdamp_char} 
\end{equation}
Note that $t_{\rm damp}$ is velocity independent.

With these reacceleration times in mind, we can now estimate the fraction of turbulent energy forcing $\tilde{\epsilon}$ that goes toward CRs. It is given by 
\begin{equation}
    f_{\rm CR} \sim \frac{\dot{E}_{\rm CR}}{\tilde{\epsilon}} \sim \frac{E_{\rm CR}}{\tilde{\epsilon} t_{\rm grow}}
\end{equation}
For example, for Kolmogorov turbulence, where $\tilde{\epsilon} \sim \rho v^{3}/L$, and for the characteristic growth time $t_{\rm grow} \sim 9/2 v_{ph} L/v^{2}$ this gives:

\begin{equation}
    f_{\rm CR} \sim \frac{E_{\rm CR}/t_{\rm grow}}{\rho v^{3}/L} \sim \rm max \left(\frac{2}{3} \mathcal{M}_{\rm ph}  \frac{P_{\rm CR}}{\rho v^{2}}, 1 \right) 
    \label{eqn:f_CR} 
\end{equation}

Note that Equation \ref{eqn:f_CR} is approximate and assumes $\tilde{\epsilon} \sim \rho v^{3}/L$, which is only true in the limit where CRs do not back-react on the flow. In general, $\tilde{\epsilon} - \rho v^3/L - E_{CR} / t_{grow} \sim 0$, and $f_{CR} \sim E_{CR}/t_{grow} / \tilde{\epsilon} \sim (E_{CR}/t_{grow}) (\rho v^3/L)^{-1} (1-f_{CR})$. This gives 

\begin{equation}
\begin{split}
    f_{CR} & \sim \frac{E_{CR}}{t_{\rm grow} (\rho v^3/L)}\left(1+\frac{E_{CR}}{t_{\rm grow} (\rho v^3/L)}\right)^{-1} \\
      & \sim  \left(\frac{2}{3} \mathcal{M}_{\rm ph}  \frac{P_{\rm CR}}{\rho v^{2}}\right) / \left(1 + \frac{2}{3} \mathcal{M}_{\rm ph}  \frac{P_{\rm CR}}{\rho v^{2}}\right)
    \label{eqn:f_CR_alt}
\end{split}
\end{equation}
which agrees with Equation \ref{eqn:f_CR} in the appropriate limits.

\begin{figure}
\centering
\includegraphics[width=0.48\textwidth]{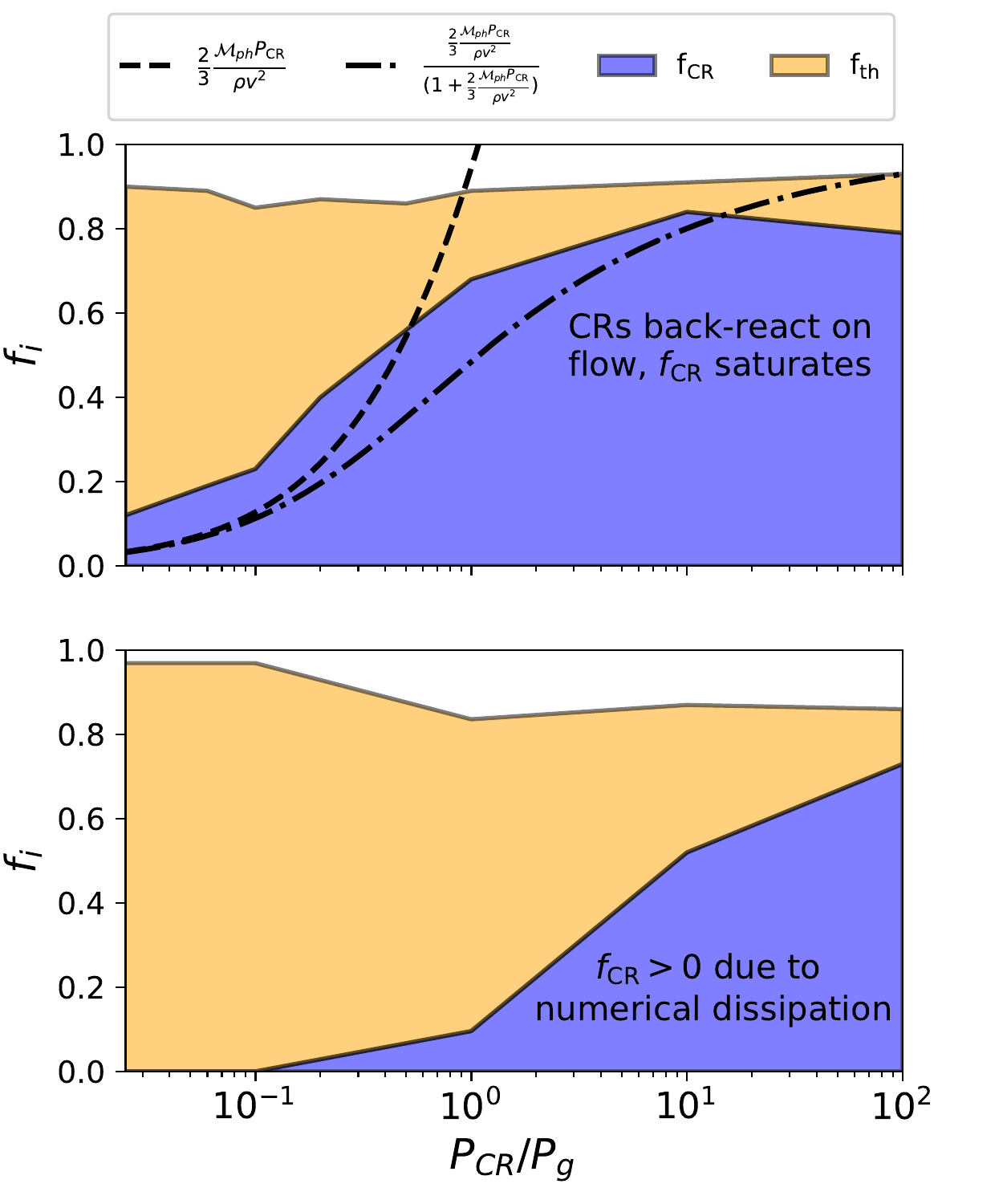}
\caption{The average CR energy gain rate and thermal energy gain rate relative to the turbulent driving rate ($f_{CR} = \dot{E}_{CR}/\tilde{\epsilon}$ and $f_{g} = \dot{E}_{g}/\tilde{\epsilon}$, respectively) for simulations without streaming, as a function of $P_{\rm CR}/P_{\rm g}$. These all are adiabatic, $\mathcal{M}_{s} \sim 0.5$ simulations on a $128^{3}$ grid, with $\beta \sim 1$. \emph{Top:} $\kappa_{||} \sim 0.15 L_{0} v_{ph}$, where CR energy gain is maximized. The dashed black curve is the analytic expectation from Equation \ref{eqn:f_CR}, showing good agreement when $P_{CR}/P_{g} < 1$, and the dash-dotted curve shows Equation \ref{eqn:f_CR_alt}, which accounts for CR back-reaction on the flow and subsequent saturation of $f_{CR}$. \emph{Bottom:} $\kappa_{||} \sim 0$. For $P_{CR} \gg P_{g}$, even $\kappa \sim 0$ leads to significant fractions of turbulent energy converted to CR energy, but this CR reacceleration is due to numerical diffusion caused by finite resolution.} 
\label{fig:fCR}
\end{figure}

\begin{figure*}
\centering
\includegraphics[width=0.94\textwidth]{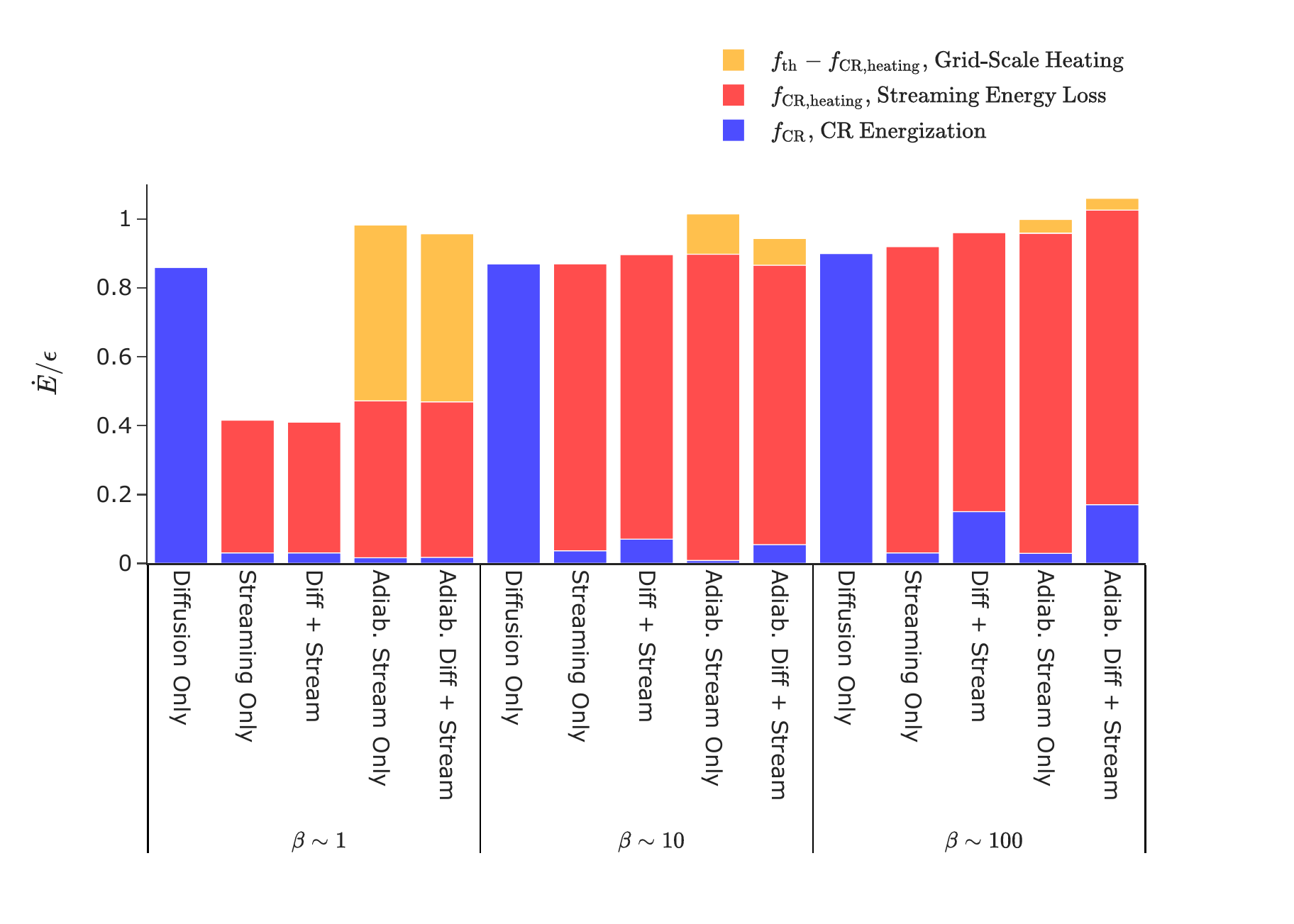}
\caption{Partitioning of input turbulent energy rate $\tilde{\epsilon}$ into three different channels: CR reacceleration $f_{CR}$, dissipation via CR collisionless heating $f_{\rm CR, heating}$ (i.e. streaming energy loss), and grid-scale heating $f_{th} - f_{\rm CR, heating}$. Without CRs, this choice of $\tilde{\epsilon}$ produces $\mathcal{M}_{\rm s} \sim 0.15$ turbulence. Each simulation here starts with $P_{CR} \sim P_{g}$ but with varying CR transport treatments, either with diffusion only (all with $\kappa \sim 0.15 v_{ph} L_0$) or diffusion plus additional streaming. For each $\beta$, the first three simulations use an isothermal equation of state, so there is no gas heating. The last two, denoted by ``Adiab.", use an adiabatic equation of state, in which case the total thermal gas heating rate is the sum of CR heating and grid-scale heating. With diffusion only, reacceleration is very efficient: most turbulent energy is soaked up by CRs. With streaming, both gas heating and CR energization are relatively inefficient in the low-$\beta$ regime, but for $\beta \sim$ 10, 100, CR heating is the dominant energy channel. Instead of turbulent energy cascading to small scales and eventually dissipating into thermal energy at the grid scale, CRs intercept this energy transfer at large scales; astoundingly, even in these subsonic flows very high fractions of turbulent energy are channelled through CRs when $P_{CR} \gsim P_{g}$. }
\label{fig:barchart_iso}
\end{figure*}

In both Equations \ref{eqn:f_CR} and \ref{eqn:f_CR_alt}, the maximum value of 1 reflects energy conservation: CRs cannot gain more energy than is injected by turbulent forcing, hence $f_{CR} \sim \frac{2}{3} \mathcal{M}_{\rm ph}  \frac{P_{\rm CR}}{\rho v^{2}}$ is only valid for $\dot{E}_{\rm CR} < \tilde{\epsilon}$. Within this regime, the fraction of kinetic energy deposited into CRs is small if $P_{\rm CR} \ll \rho v^{2}$, in which case most energy is deposited in the thermal gas; however, for higher $P_{\rm CR}$, the fraction increases and can become quite substantial at close to equipartition values.

Figure \ref{fig:fCR} compares this expectation to simulations and is one of the key results of this paper. The y-axis shows the partitioning of the input energy rate into CRs ($f_{CR} = \dot{E}_{\rm CR}/\tilde{\epsilon}$) and thermal energy ($f_{th} = \dot{E}_{\rm th}/\tilde{\epsilon}$) for varying $P_{\rm CR}/P_{\rm g}$, keeping $\tilde{\epsilon}$ fixed, for purely diffusive CRs. Unlike our previous simulations, which all used an isothermal equation of state, these simulations have an adiabatic equation of state, which makes it easier to confirm energy conservation. Together, the contributions to $\dot{E}_{\rm CR}$ and $\dot{E}_{\rm th}$ sum to $\sim 80-90\%$ of the driving rate, with the rest going towards small magnetic and kinetic energy increases. The top and bottom panels show simulations each without streaming and with $\kappa = 0.15 L_{0} v_{ph}$ and $\kappa = 0$, respectively. For $P_{CR}/P_{g} < 1$, $f_{CR}$ follows the expectation from Equation \ref{eqn:f_CR} (shown as a black dashed line) quite well, an indication that turbulent reacceleration is diverting the driving energy to CRs at the expense of thermal gas heating. Similar simulations with\footnote{In practice, $\kappa$ has a non-zero value because of numerical diffusion, but here this has little impact up until $P_{CR} \gg P_{g}$.} $\kappa = 0$ show far lower $f_{CR}$, again revealing the dependence of reacceleration on diffusion coefficient. Note that while we previously only tested analytic expectations for the growth time $t_{\rm grow}$ (on which Equation \ref{eqn:f_CR} depends) when the gas is isothermal in Paper I, they continue to hold when the gas is adiabatic. 

As $P_{CR}/P_{g}$ increases, $f_{\rm CR}$ deviates from the analytic expression in Equation \ref{eqn:f_CR}; $f_{\rm CR}$ increases more slowly towards the asymptotic bound $f_{\rm CR} \sim 1$ than in our ansatz. Nonetheless, for $P_{CR}/P_{g} \gsim 1$, what immediately stands out is the large fraction of energy diverted to CRs, with $f_{CR}$ as large as 0.8 when $P_{CR}/P_{g} > 1$. These large values of $f_{CR}$ clearly come at the expense of thermal heating\footnote{Since we enforce purely compressive driving, magnetic field amplification is very weak, and $f_{\rm CR} + f_{\rm th} \approx 1$ for an adiabatic setup.}, with $f_{\rm th}$ decreasing from $f_{\rm th} \approx 1$ when $P_{CR} \ll P_{g}$ to $f_{\rm th} < 0.2$ when $P_{CR} > P_{g}$.

In the above, purely diffusive case, turbulent energy directly accelerates CRs. When streaming is included, energy is also lost to collisionless heating at a rate $H = v_{A} \cdot \nabla P_{CR}$. In Fig \ref{fig:barchart_iso}, we quantify the partitioning of turbulent kinetic energy into direct acceleration of CRs ($f_{\rm CR}$) and gas heating ($f_{\rm th}$) in simulations with fixed $\tilde{\epsilon}$ producing undamped $\mathcal{M}_{\rm s} \sim 0.15$. We distinguish between collisionless heating by CRs $f_{\rm CR,heating}$ (red bars), and heating due to turbulence which cascades down to the grid scale and dissipates $f_{th} - f_{\rm CR,heating}$ (orange bars). Note that, in all cases (see e.g. the adiabatic $\beta = 10$, 100 simulations), the sum of $f_{\rm CR}$, $f_{\rm CR,heating}$, and $f_{th} - f_{\rm CR,heating}$ can be slightly greater or slightly lower than 1; we average each dE/dt over the final 1/4 of the simulation snapshots, and during this time interval of fully developed turbulence, kinetic and magnetic energy can, on average, be slightly decreasing or slightly increasing. For that reason, the sum of all bars shown for each simulation in Figure \ref{fig:barchart_iso} lands between 0.95 and 1.05 of the input driving rate.

When streaming is included, $f_{\rm CR}$ is a small and weakly increasing function of $\beta$, consistent with Paper I and evident in Figure \ref{fig:barchart_iso}. Here, we fix the initial state to have $P_{CR} \sim P_{g}$ for each simulation and quantify the CR energy gain rate as we did in Figure \ref{fig:fCR}. Despite the fact that a negligible fraction of energy $f_{CR}$ ends up in CRs, the latter nonetheless have a strong impact on the turbulent cascade. In MHD simulations, turbulence cascades to grid scales where numerical diffusion dominates\footnote{In high resolution simulations with explicit viscosity, it would instead cascade to the viscous scale.} and subsequently dissipates, heating the gas. Thus, $f_{th}$ is a good barometer of how much kinetic energy flux makes it to the dissipation scale; however, that is not the case with turbulence modified by streaming CRs. In the adiabatic streaming simulations quantified in Figure \ref{fig:barchart_iso}, the total amount of gas heating is a weak function of $\beta$, but actually much of that heating is done by CR streaming energy loss instead of classical small-scale dissipation. 
Only $\sim 60 \%$ (for $\beta \sim 1$) to $<10 \%$ (for $\beta \sim 10$, $100$) of the driving energy makes it to the grid scale, with the remaining energy channeled through CRs.

Note that in our estimate of $t_{\rm damp}$ (Equation \ref{eq:tdamp_char}), we have {\it not} included the effects of CR streaming on $t_{\rm grow}$. If we did, $t_{\rm damp}$ would be substantially longer in low $\beta$ environments.  However, as we have seen, this is incorrect. When CR streaming is present, the kinetic energy of compressive motions is still absorbed by CRs at large scales. This energy is subsequently returned to the gas in the form of heat via CR streaming, and so streaming impedes the secular growth of CR energy, resulting in the lower growth times explored in Paper I. However, diversion of kinetic energy away from the turbulent cascade and damping of compressive motions still happens at a similar rate, even at low $\beta$ (Figure \ref{fig:barchart_iso}). CR streaming provides an avenue for gas motions to quickly dissipate in the form of heat without going through the turbulent cascade. In this case, CRs can be thought of as providing an unusual form of viscosity.

To summarize: once $P_{\rm c}/P_{\rm g} \gsim 1$, and for $\beta \gsim 10$, our simulations show that the energy input in turbulent driving appears to be almost completely diverted to CRs, with only $\sim 10\%$ remaining which cascades down to grid scales. This is irrespective of whether streaming is absent (in which case CRs store the energy) or present (in which case CRs thermalize a significant fraction via collisionless heating). This is astonishing efficiency, considering that strong shocks convert at best $\sim 10-30\%$ of kinetic energy to CRs. For $\beta \sim 1$, the fraction of energy routed through CRs is slightly lower, $\sim 80\%$ in the diffusion only case, and $\sim 50\%$ with both CR streaming and diffusion. We now turn to some implications of this finding.

\section{Cosmic Ray Imprints on Kinetic Energy Spectra}
\label{sec:spectra}

In certain regimes, CRs are clearly an important energy sink for fluid motions. When turbulent energy is diverted to the CR population, it either

\begin{enumerate}
\item Directly accelerates CRs through non-resonant reacceleration 
\item (If CR streaming is significant) Heats the gas at scales $l_{\rm CR} \gg l_{\rm diss}$ through collisionless energy transfer by self-confined CRs (streaming energy loss), where $l_{\rm diss}$ is the Kolmogorov dissipation scale. 
\end{enumerate}
In either case, energy that originally would have cascaded to small scales is siphoned out of the turbulent cascade, and it is interesting to ask what imprint this might have on the kinetic energy spectrum. In this section, we first focus on the effects of purely diffusive CRs, leaving an initial exploration of streaming CR transport, the effects of which are less straightforward and deserve future follow-up, to \S\ref{sec:streaming}. We will first explore CR modifications to Kolmogorov and Kraichnan spectra analytically and discuss astrophysical regimes where spectra could be heavily modified. Of the compressible MHD modes, it is thought that slow modes have a Kolmogorov spectrum ($E(k) \propto k^{-5/3}$) and fast modes have a Kraichnan spectrum ($E(k) \propto k^{-3/2}$) \citep{cho2003}, though this is still debated. In our simulations, compressive forcing gives rise to something intermediate between Kraichnan and Burgers turbulence ($E(k) \propto k^{-2}$), and we will see that CR damping also has noticeable effects in this regime.

\subsection{Analytic Theory}
\label{sec:damping_analytics}

We can solve for the turbulent power spectrum by solving the dynamic equation \citep{landau87}. If we consider a turbulent energy injection rate $\epsilon$ injected at some outer scale $L = k_{\rm L}^{-1}$ (where $\epsilon \sim v_{l}^{2}/t_{\rm cascade}\sim$ const in the absence of damping, and $t_{\rm cascade}$ depends on the form of turbulence), then in steady state the combined effects of the cascade to smaller scales and damping must balance injection: 
\begin{equation}
    \epsilon \ \delta (k-k_{\rm L}) = \frac{\partial}{\partial k} F(k) + \Gamma(k) E(k) 
    \label{eqn:dynamic}
\end{equation}
where $E(k)$ is the power spectrum of turbulence, $F(k)$ is the turbulent cascade flux in k-space, and $\Gamma(k) \sim t_{\rm damp}^{-1}$ is the damping rate. While Equation \ref{eqn:dynamic} makes no assumption on the turbulent spectrum or the damping rate, we now must adopt choices for each. First, Equation 4 in Paper I describes the CR reacceleration rate $t_{\rm grow}^{-1}$ from an ensemble of waves across many scales; however, to assess the impact of CRs on turbulence at a given scale, we need to consider just the amount of energy that CRs sap from individual eddies of scale $l$. Assuming we are in the fast transport regime ($\kappa > v_{ph} l$),  $t_{\rm grow} \sim \kappa/v_{l}^{2}$, hence, $t_{\rm damp} \sim \rho v_{l}^{2}t_{\rm grow}/P_{\rm CR} \sim \kappa/c_{c}^{2}$. $\Gamma(k) = t_{\rm damp}^{-1}$ is then scale-independent. In the slow transport regime, $t_{\rm grow} \sim v_{ph}^{2} l^{2} /(v_{l}^{2} \kappa)$ and $t_{\rm damp} \sim \rho v_{ph}^{2} l^{2} /(P_{\rm CR} \kappa)$. Because the latter is scale-dependent, we'll make the simplifying assumption that diffusion is fast, such that $t_{\rm damp}$ is scale-independent. This is not unreasonable, especially at small scales, because for a given $\kappa_{||}$, transport across smaller and smaller scales is increasingly in the fast regime. As we'll see, our simulations display similar behavior to our following analytics that assume fast diffusion.

While the above $\Gamma(k)$ is scale-independent and therefore makes no assumption on cascade physics, the cascade flux $F(k)$ depends on the type of turbulence: for Kolmogorov turbulence, $F(k) \sim [E(k)]^{3/2} k^{5/2}$, while for isotropic Kraichnan turbulence, $F(k) \sim k^3 [E(k)]^2/v_{\rm ph}$. In the absence of damping ($\Gamma(k)=0$), integrating both sides of Equation \ref{eqn:dynamic} with respect to $k$ gives $E(k) \sim \epsilon^{2/3} k^{-5/3}$ and $E(k) \sim (\epsilon v_{\rm ph})^{1/2} k^{-3/2}$, the power spectra for Kolmogorov and Kraichnan turbulence respectively.

\begin{figure*}
\centering
\includegraphics[width=0.95\textwidth]{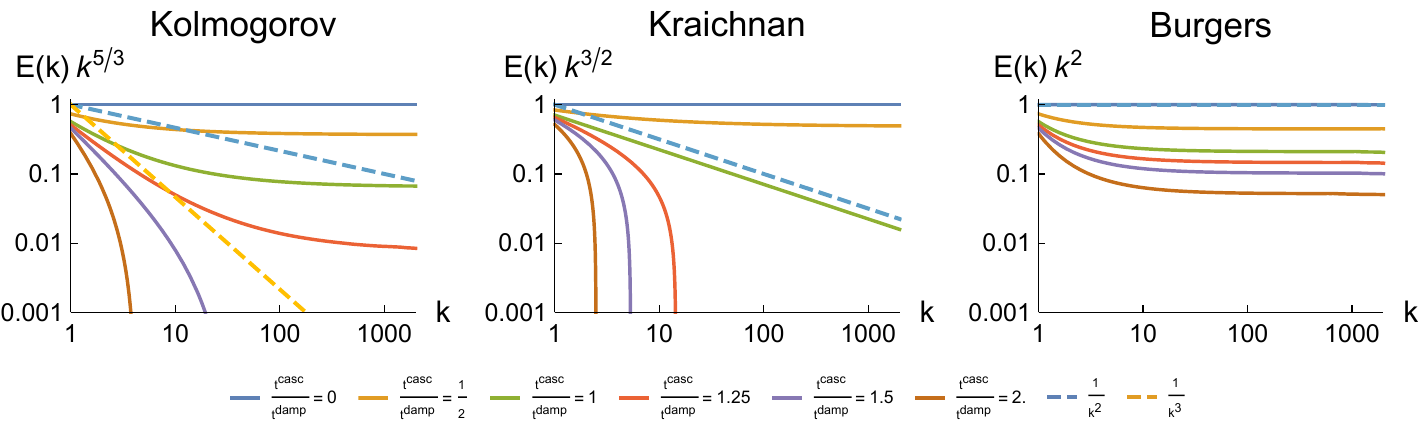}
\caption{Modified kinetic energy spectra for a Kolmogorov (left), Kraichnan (middle), and Burgers (right) cascade with varying levels of CR damping, all with $v_{ph} = 2 v_{0}$. E(k) is in units of the outer-scale, undamped kinetic energy, where k denotes the wavenumber. Different line colors denote different ratios of the cascade time to the damping time, showing that if damping becomes competitive, the outer scale velocity decreases, and the slope of the spectrum steepens. Dashed lines show $E(k) = k^{-2}$ and $E(k) = k^{-3}$ for comparison. For $t_{\rm cascade}/t_{\rm damp} \gtrapprox 1.5$, 1 for Kolmogorov and Kraichnan, respectively, the cascade sharply cuts off at progressively smaller k. For smaller $t_{\rm cascade}/t_{\rm damp}$, CRs damp fluctuations, but the cascade returns to its normal scaling at large k. For Burgers turbulence, which is not a genuine cascade, there can be an appreciable decrease in power at small k, but at high k, the spectrum recovers a $k^{-2}$ slope.} 
\label{fig:dynamic}
\end{figure*}

The first and second terms on the right hand side of Equation \ref{eqn:dynamic} have units of $v^{2}/k \times (t_{\rm cascade}^{-1}, t_{\rm damp}^{-1})$ respectively. In Fig. \ref{fig:dynamic}, we solve Equation \ref{eqn:dynamic} for various values of $t_{\rm damp}/t_{\rm cascade}$. It is easy to understand the asymptotic behavior. When $t_{\rm cascade} \ll t_{\rm damp}$, the first term on the RHS dominates: injected energy cascades before it can damp, and we obtain the usual Kolmogorov/Kraichnan power spectra. On the other hand, if $t_{\rm damp} \ll t_{\rm cascade}$, then the second term on the RHS dominates, which gives $\epsilon \sim \Gamma \int E(k) dk \sim \Gamma v^2$, or
\begin{equation}
    v^{2} \sim \epsilon t_{\rm damp} \sim v_0^{2} \left( \frac{t_{\rm damp}}{t_{\rm cascade}} \right) 
    \label{eqn:veldrop}
\end{equation}
where $v_0^2$ and $t_{\rm cascade}$ are the velocity and cascade time at the outer scale in the absence of damping; for a given energy forcing $\epsilon$, the velocity at the outer scale is reduced. However, since $t_{\rm damp} \gsim t_{\rm inject}$, the damping time cannot be made arbitrarily small. We discuss this further in \S\ref{sec:tcascade}.

To understand the behavior at smaller scales, note that the cascade time is scale-dependent, while for non-resonant CR acceleration, $t_{\rm damp}$ is independent of scale. We are accustomed to thinking of the cascade time decreasing towards small scales (for instance, $t_{\rm cascade} \propto l^{2/3}, l^{1/2}$ for undamped Kolmogorov, Kraichnan turbulence respectively). However, damping changes the scale dependence of velocity, further reducing velocities at small scales, and thus increasing cascade times at these scales. If $t_{\rm cascade}/t_{\rm damp}$ still decreases towards small scales, then the cascade eventually takes over and the spectrum rebounds from damping. However, if $t_{\rm cascade}/t_{\rm damp}$ instead increases towards small scales, then damping becomes increasingly dominant and the spectrum will cut off precipitously. Since $t_{\rm damp}$ is independent of $k$, what matters is the scale dependence of $t_{\rm cascade}$. 

From Equation \ref{eqn:dynamic}, the cascade time can be written as: 
\begin{eqnarray} 
t_{\rm cascade} \sim \frac{k E(k)}{F(k)} &\sim& \frac{1}{[k^{3} E(k)]^{1/2}} \ {\rm (Kolmogorov)} \\ 
&\sim& \frac{v_{\rm ph}}{k^{2} E(k)} \ {\rm (Kraichnan)} 
\label{eq:tcascade_k} 
\end{eqnarray} 
where we have used $F(k) \sim [E(k)]^{3/2} k^{5/2}$, $F(k) \sim k^3 [E(k)]^2/v_{\rm ph}$ for Kolmogorov and Kraichnan turbulence respectively. When damping operates, $E(k)$ will steepen from standard Kolmogorov/Kraichnan spectra. From Equation \ref{eq:tcascade_k}, we see that for a power spectrum $E(k) \propto k^{-\alpha}$, $t_{\rm cascade}$ {\it increases} with $k$ for $\alpha \gsim 3$ (Kolmogorov), $\alpha \gsim 2$ (Kraichnan). The steepening of the power spectrum slope is controlled by the relative strength of damping, i.e. $t_{\rm cascade}/t_{\rm damp}$ at large scales. If this is sufficiently large, it produces a power spectrum with a slope steeper than the critical value, and we have a runaway: $t_{\rm cascade}/t_{\rm damp}$ continually increases towards small scales, producing a rapid cutoff in the velocity power spectrum. However, if the initial value of $t_{\rm cascade}/t_{\rm damp}$ produces a power spectrum with an index shallower than the critical slope, then damping initially `takes a bite' out of the turbulent cascade, but $t_{\rm cascade}/t_{\rm damp}$ decreases towards small scales, until damping becomes negligible, the original cascade dominates and the spectrum recovers its original undamped power law slope. 

We clearly see confirmation of this bifurcation in small scale damping in Fig. \ref{fig:dynamic}. We see that we require $t_{\rm cascade}/t_{\rm damp} \gsim 1.5$ at the outer scale for critical damping in a Kolmogorov cascade (so that the power spectrum steepens beyond $E(k) \propto k^{-3}$), or $t_{\rm cascade}/t_{\rm damp} \gsim 1$ at the outer scale for critical damping in a Kraichnan cascade (so that the power spectrum steepens beyond $E(k) \propto k^{-2}$). Indeed, $t_{\rm cascade}/t_{\rm damp} \sim 1$ causes a perfect transformation of the Kraichnan spectrum from a $E(k) \propto k^{-3/2}$ spectrum to a Burgers-like $E(k) \propto k^{-2}$ spectrum.

This bifurcation in the existence of small scale turbulence is important, so we restate it in simpler terms. Damping can change the slope of the velocity power spectrum $E(k) \propto k^{-\alpha}$, and hence the scale dependence of velocity $v(k) \propto k^{(1-\alpha)/2}$ (using $v^{2} \sim k E(k)$), but it does not change the physics of the turbulent cascade. The latter can be encapsulated in the form of cascade times $t_{\rm cascade} \sim l/v_{\rm l}$ (Kolmogorov), $t_{\rm cascade} \sim l v_{\rm ph}/v_{\rm l}^{2}$ (Kraichnan). Using $v(k) \propto k^{(1-\alpha)/2}$, these relations imply $t_{\rm cascade} \propto k^{(\alpha-3)/2}$ (Kolmogorov), and $t_{\rm cascade} \propto k^{\alpha-2}$ (Kraichnan), which gives critical slopes $\alpha=3,2$ respectively, in line with our previous arguments.  
The scale dependence of $t_{\rm cascade}$ determines if turbulence is completely damped at small scales, or recovers with the original (undamped) power-law scaling.

The right panel of Figure \ref{fig:dynamic} shows modified ``Burgers" spectra where we've solved Equation \ref{eqn:dynamic} with $F(k) \sim k^{2}E(k)$. In this case, even when $t_{\rm cascade}/t_{\rm damp} > 1$, the modified kinetic energy spectra never show cut-offs, instead always converging to a $k^{-2}$ spectrum at high k, but there is a substantial decrease in small-scale power compared to the undamped case. We briefly note that Equation \ref{eqn:dynamic} does not really apply to Burgers turbulence $E(k) \propto k^{-2}$, which is not a genuine turbulent cascade, but rather an instantaneous jump from large to small scales via shocks which arise from non-linear steepening. However, Ptuskin damping creates friction which can balance non-linear steepening and prevent shock formation. We can see this by examining Burgers' equation in the presence of Ptuskin damping: 
\begin{equation}
\frac{\partial v}{\partial t} + v \cdot \nabla v = - \Gamma v
\label{eq:burgers} 
\end{equation} 
For $\Gamma > \nabla v$, the damping term exceeds the non-linear term, so that damping exceeds non-linear steepening when the nonlinear time $t_{\rm NL} \sim L/v > t_{\rm damp}$. The outcome of this is uncertain. Figure \ref{fig:dynamic} suggests that wave amplitudes will be most significantly damped at low k, after which steepening still occurs but with reduced amplitude. In any case, $t_{\rm NL}/t_{\rm damp}$ potentially plays a similar role to $t_{\rm cascade}/t_{\rm damp}$, and as such, we will use $t_{\rm NL}$ as a proxy for $t_{\rm cascade}$ in our simulation analysis (\S \ref{sec:damping-sims}).

\subsection{What is $t_{\rm cascade} / t_{\rm damp}$?}
\label{sec:tcascade} 

The results of the previous section show that the ratio $t_{\rm cascade}/t_{\rm damp}$ is the critical parameter determining the efficacy of small scale damping, and that there is a critical value ($t_{\rm cascade}/t_{\rm damp} \gsim 1.5,1$ for Kolmogorov and Kraichnan turbulence, respectively) such that the turbulence spectrum will show a cutoff. Here, we investigate the conditions under which these thresholds may be crossed.

We have previously argued from energy conservation that $\dot{E}_{\rm CR} \lsim \tilde{\epsilon}$ in steady state, hence $t_{\rm damp} \ge t_{\rm inject} \sim \rho v^{2}/\tilde{\epsilon} \sim L/v$, the timescale on which kinetic energy is injected. In Appendix \ref{appendix:turb}, we confirm this expectation and also show how various scalings, such as $\delta \rho/\rho, \delta v/v$, can be understood as a function of $P_{\rm CR}/P_g$, or  $v/c_{\rm s}, v/v_{\rm ph}$.

When does $t_{\rm damp}$ reach the minimal value of $t_{\rm inject} \sim L/v$, so that almost all of the injected kinetic energy is directly dissipated in cosmic rays? Equating the first and second terms in brackets in Equation \ref{eq:tdamp_char}, $t_{\rm damp} \sim t_{\rm inject}$ when: 
\begin{equation} 
\mathcal{M}_{\rm ph} \lsim 
\left( \frac{P_c}{P_{\rm tot}} \right)
\label{eq:tdamp_min}
\end{equation} 
Equation \ref{eq:tdamp_min} is only an order of magnitude estimate; the exact threshold must come from numerical simulations. Nonetheless, it illustrates the relevant physics: damping saturates when the turbulent Mach number is small and the CR energy density is high.

If $t_{\rm damp}$ reaches its minimal value of $t_{\rm inject} \sim L/v$, then: 
\begin{equation} 
\begin{split}
    \frac{t_{\rm cascade}}{t_{\rm damp}} & \sim 1 \quad {\rm (Kolmogorov)} \\ 
    & \sim \frac{1}{\mathcal{M}_{\rm ph}} \quad {\rm (Kraichnan)}
\end{split}
\label{eq:ratio_t_cascade_damp} 
\end{equation}
From Fig \ref{fig:dynamic}, we see that it is unclear whether damping will be strong enough to enforce a small scale cutoff in a Kolmogorov cascade (which requires ${t_{\rm cascade}}/{t_{\rm damp}} \gsim 1.5$), but any subsonic turbulence in a Kraichnan cascade which satisfies Equation \ref{eq:tdamp_min} will automatically have ${t_{\rm cascade}}/{t_{\rm damp}} \gsim 1$), the threshold for critical damping there. The increase in ${t_{\rm cascade}}/{t_{\rm damp}}$ is not due to a decrease in the damping time (which has a floor at $t_{\rm inject}$), but rather the increased cascade time in MHD turbulence. Longer cascade times are associated with wave turbulence, where wave-wave interactions produce non-linearities which eventually cause turbulence to cascade \citep{nazarenko11-book}. Other forms of wave turbulence can be present, for instance, in systems with strong stratification \citep{wang22} or rotation. 

Note that even if the threshold for critical damping (i.e. exponential suppression of small-scale power) is not met, Figure \ref{fig:dynamic} shows that the damping of gas motions can still be significant.

\subsection{Simulations}
\label{sec:damping-sims}

\begin{figure*}
\centering
\includegraphics[width=0.9\textwidth]{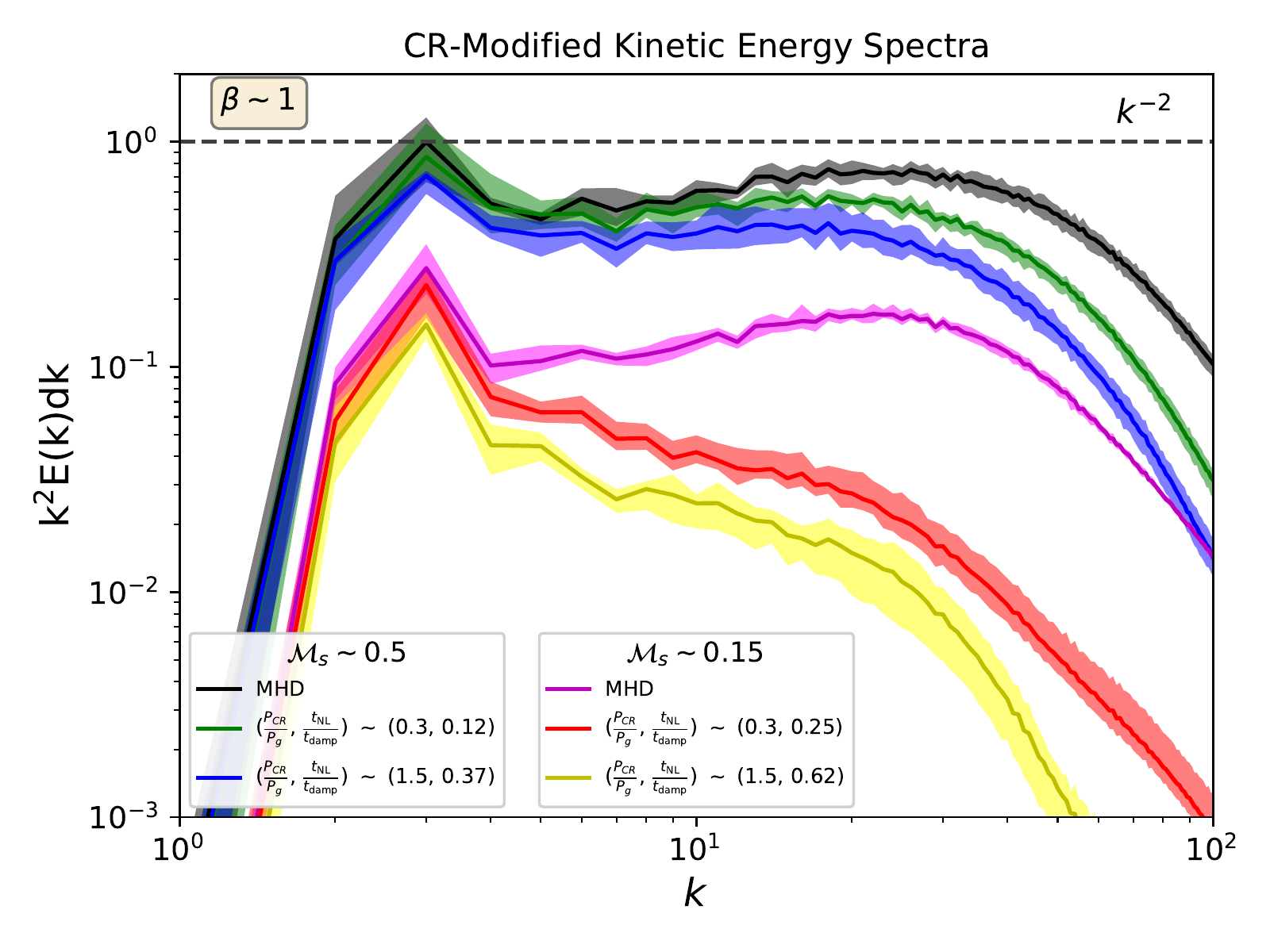}
\caption{The turbulent kinetic energy spectrum, multiplied by $k^{2}$ for a set of $\mathcal{M}_{s} \sim 0.5$ and $\mathcal{M}_{s} \sim 0.15$ diffusion-only simulations, keeping $\kappa = 0.15 L_{0} v_{ph}$ and $\beta \sim 1$, as we vary $P_{CR}/P_{g}$. Spectra are normalized to the k=3 mode for the $\mathcal{M}_{s} \sim 0.5$ MHD run. Ratios of the \emph{outer scale} nonlinear time to damping time, calculated with $t_{\rm damp} = <\rho v^{2}/\dot{P}_{CR}>$ and $t_{\rm NL} = L_0/v$, are also denoted. Points show the energy in each k-bin averaged over 10 outputs at late times, when turbulence is fully developed, while shaded regions show the minima and maxima during those time periods. Each simulation was run on a $512^{3}$ grid. While MHD runs produce overall spectra shallower than $k^{-2}$, CRs damp fluctuations, slightly decreasing the power in low k modes while steepening the spectra at high k. } 
\label{fig:CR_Turb_Modify}
\end{figure*}

The results of \S\ref{sec:damping_analytics}, \ref{sec:tcascade} are useful for guiding expectations and driving intuition. Nonetheless, given the complex non-linearities, they require validation by numerical simulation -- a difficult task, given the limited inertial range of standard resolution simulations. We now present a set of simulations which, to our knowledge, are the first CR hydrodynamics simulations specifically probing CR influence on turbulent kinetic energy spectra. While a more complete set of simulations with different driving modes and higher resolution awaits, we already see that CRs suppress small-scale fluctuations. 

We focus first on the case where Ptuskin damping is maximized, running a series of diffusion-only simulations near the CR energy gain `sweet spot' $\kappa \sim 0.15 L_0 v_{\rm ph}$, where $v_{\rm ph}^2 \sim (P_c+P_g+P_B)/\rho$. We vary the input driving rate $\tilde{\epsilon}$ by an order of magnitude to create turbulence with undamped $\mathcal{M}_{\rm s} \sim 0.5$ and $\mathcal{M}_{\rm s} \sim 0.15$ , where $\mathcal{M}_{\rm s} = v/c_s$ and $c_s \sim \sqrt{P_g/\rho}$ is the gas sound speed (thus, $\mathcal{M}_{\rm ph}=v/v_{\rm ph}$ decreases as $P_c/P_g$ increases). Plasma beta, $\beta = P_{g}/P_{B}$, are denoted in each figure and represent rough values for the presented suite of simulations; while each simulation starts with the same $\beta$, the saturated value of $\beta$ changes by a small amount depending on whether CRs are present, what CR transport model is assumed, etc. 

Figure \ref{fig:CR_Turb_Modify} shows simulation kinetic energy spectra for both $\mathcal{M}_{\rm s} \sim 0.5$ and $\mathcal{M}_{\rm s} \sim 0.15$ simulation sets, each normalized by the k = 3 mode power for the $\mathcal{M}_{\rm s} \sim 0.5$ MHD-only simulation. Different colors denote different initial $P_{\rm CR}/P_{g}$, ranging from $0$ to $1.5$. Points denote average kinetic energies, and the shaded regions denote the minimum and maximum kinetic energies taken over 10 snapshots at late times when we see converged spectra, typically between 8 and 10 eddy turnover times after the simulation starts (see Appendix for more about time convergence). Importantly, we note that the inertial ranges in our MHD-only simulations display something between a Kraichnan ($k^{-3/2}$) and a Burgers-like ($k^{-2}$) spectrum, with significant power at high k due to the generation of solenoidal modes rather than fast modes despite our purely compressive forcing (see Section \ref{sec:comp_vs_sol}). The $k^{-2}$ compressive component we find is frequently seen in hydrodynamic simulations with compressive driving, due to non-linear steepening (e.g., \citealt{miniati15}). Thus, the analytic models of \S\ref{sec:damping_analytics},\ref{sec:tcascade} where we assume a Kraichnan spectrum do not exactly apply. Nonetheless, we can look for qualitative agreement. 

\begin{figure*}
\centering
\includegraphics[width=0.96\textwidth]{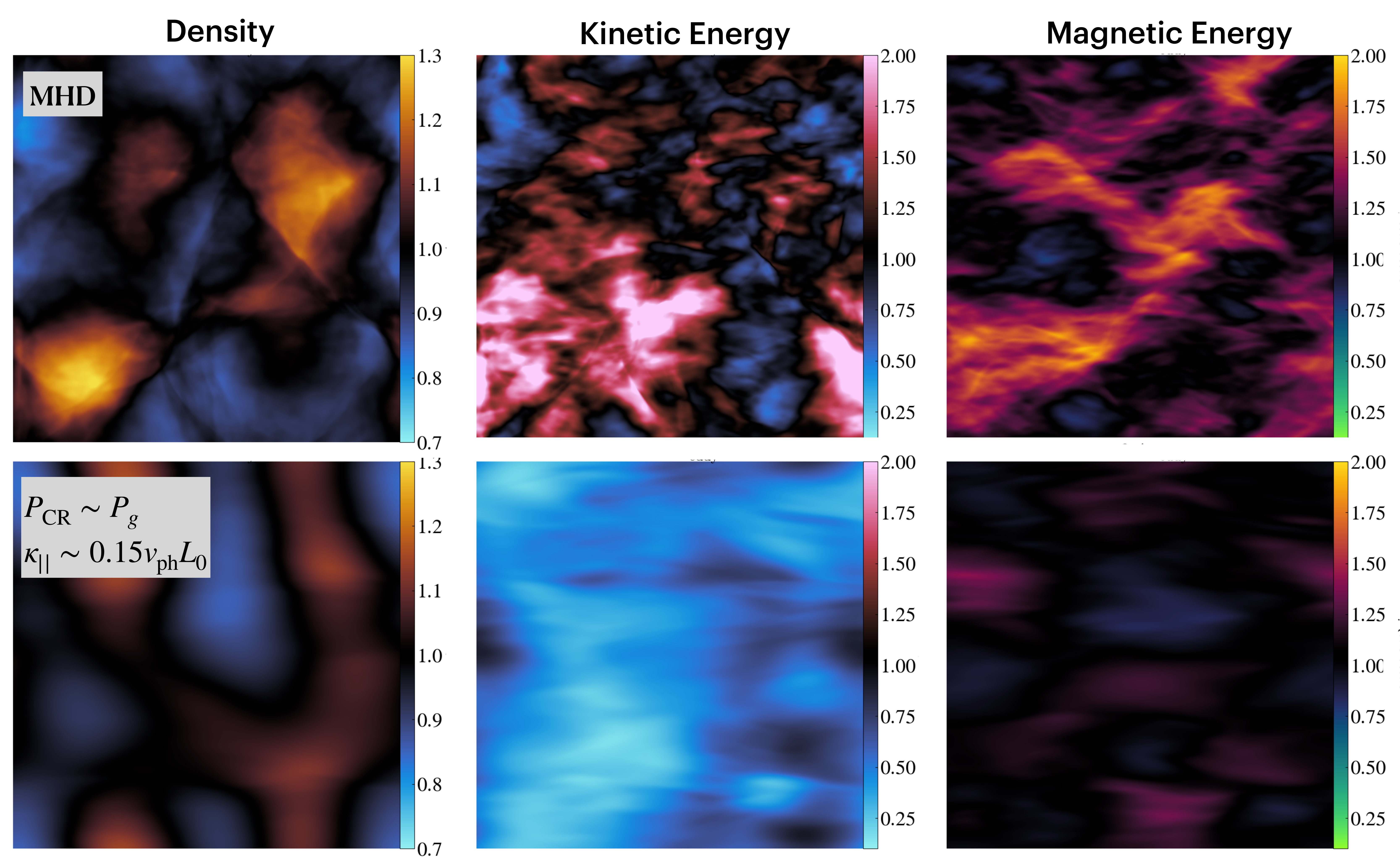}
\caption{Projections perpendicular to the initial magnetic field direction of density (left column), kinetic energy (middle column), and magnetic energy (right column) after $\sim 10$ eddy turnover times, normalized by their average values in the MHD only case. \emph{Top:} MHD-only simulations with $\beta \sim 10$ and $\mathcal{M}_{\rm s} \sim 0.15$. \emph{Bottom:} Simulations with the same $\beta$ and forcing rate, but with $P_{CR} \sim P_{g}$ and diffusive CR transport. Density, velocity, and magnetic fluctuations are all suppressed compared to the MHD case.}
\label{fig:fluctuations_pureDiffusion}
\end{figure*}

We can use the non-linear steepening time as a proxy for the cascade time: $t_{\rm casc} \sim t_{\rm NL} \sim L_0/v_{L}$, where $v_{L}$ is the outer-scale velocity. Ratios of cascade time to damping time, calculated with $t_{\rm damp} = <\rho v^{2}/\dot{P}_{CR}>$, are noted in the legend.  The trend agrees at least qualitatively with Figure \ref{fig:dynamic}. Power both at large and small scales is decreased when $P_{CR} \ge P_{g}$, consistent with mild Ptuskin damping when $t_{\rm cascade} \sim t_{\rm damp}$. As $t_{\rm cascade} / t_{\rm damp}$ increases, the spectrum deviates further and further from the MHD case. For example, the $t_{\rm cascade} / t_{\rm damp} \sim 0.62$, $\mathcal{M}_{\rm s} \sim 0.15$ simulation has between 10 and 100 times less power in high-k modes than the MHD run. Projections of density, kinetic energy, and magnetic energy for these $\mathcal{M}_{\rm s} \sim 0.15$, $\beta \sim 10$ simulations vary quite obviously, as seen in Figure \ref{fig:fluctuations_pureDiffusion}, with fluctuations clearly damped in the $P_{CR} \sim P_{g}$ case (bottom row) compared to the MHD case (top row). Higher Mach number simulations appear to show damping, as well, but the effect is less obvious. This is in line with expectations from our previous discussion that $t_{\rm cascade}/t_{\rm damp}$ is maximized for smaller values of stirring velocity.

\begin{figure*}
\centering
\includegraphics[width=0.95\textwidth]{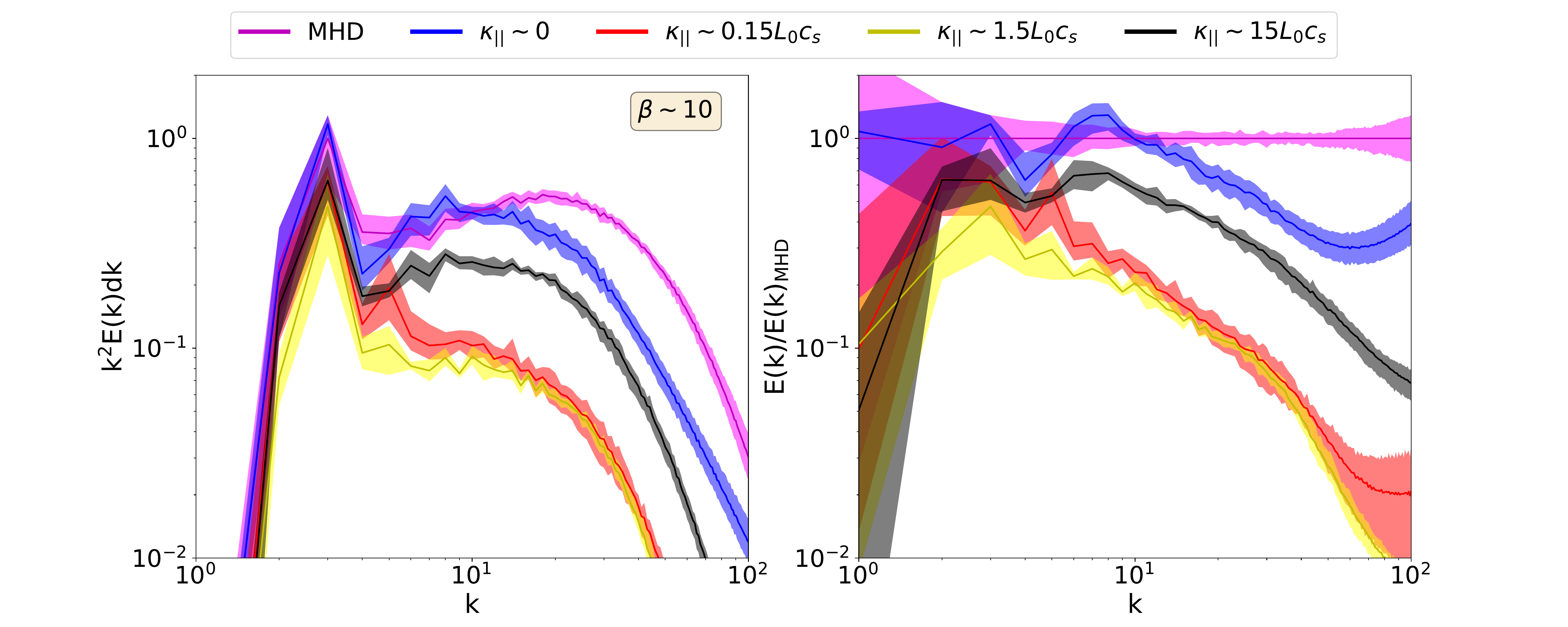}
\caption{Kinetic energy spectra of $512^{3}$ simulations when $P_{CR} \sim P_{g}$ but varying the diffusion coefficient from $\kappa_{||} \sim 0$ (where the only diffusivity is numerical) to the most efficient reacceleration regime ($\kappa_{||} \sim 0.15 v_{ph} L_0$ and $\kappa_{||} \sim 1.5 v_{ph} L_0$) to the fast diffusion regime ($\kappa_{||} \sim 15 v_{ph} L_0$). Note how the power spectrum is somewhat different for the two fluid system even in the absence of CR transport, presumably because of changes to the phase velocity and other adiabatic properties, but deviations from the MHD spectrum are mild compared to simulations with added diffusion. \emph{Left:} $k^{2}E(k)dk$ normalized by the $k = 3$ MHD value. \emph{Right:} Ratio of each spectrum to the MHD spectrum. Note how diffusion introduces a characteristic scale $l_{CR}$ where the kinetic energy is reduced: in the fast diffusion limit, $l_{CR} > L_0$, and the outer-scale kinetic energy drops significantly while the rest of the spectrum retains the same shape as the MHD case. Going to smaller $\kappa_{||}$, overall changes are more drastic because reacceleration is more efficient but also the scale where the spectrum cuts off most dramatically shifts to $l_{CR} < L_0$.} 
\label{fig:KE_spectra_varyKappa}
\end{figure*}

While our analytic predictions and preliminary simulations suggest that Ptuskin damping could play a role in suppressing the compressible turbulent cascade at small scales, it may appear hazardous to draw conclusions based on moderate resolution simulations with limited inertial range. We therefore refer the reader to Figure \ref{fig:KE_spectra_varyKappa}, which shows kinetic energy spectra for simulations on a $512^{3}$ grid, each with initial $\beta \sim 10$ but diffusion coefficients varying between $\kappa_{||} \sim (0 - 15) L_{0} c_{s}$. Clearly, the strongest damping effect occurs when $\kappa_{||}$ is near the sweet-spot ($\kappa_{||} \sim (0.15 - 1.5) L_0 c_{s}$), and the effect diminishes as $\kappa_{||}$ increases. Maybe most importantly, significant spectral changes do not occur in the absence of CR transport ($\kappa_{||} \sim 0$), suggesting that numerical diffusion plays a negligible role.

We also refer the reader back to \S \ref{sect:CR-diversion} and Figure \ref{fig:fCR}, where we presented a separate, more robust diagnostic of the suppression of the turbulent cascade by Ptuskin damping: via the heating of adiabatic gas. In hydrodynamic simulations of adiabatic gas, we have found that $\dot{E}_{\rm gas} \rightarrow \tilde{\epsilon}$, as it should. However, in adiabatic simulations with CRs, we have found $\dot{E}_{\rm gas} \rightarrow 0$, while $\dot{E}_{\rm CR} \rightarrow \tilde{\epsilon}$, i.e. almost all of the turbulent energy is absorbed by the CRs (see Figure \ref{fig:fCR}). Furthermore, all of this energy is absorbed at large scales, which are well resolved. The shift to CRs receiving almost all the energy of the turbulent cascade is genuine turbulent acceleration, not due to numerical diffusion in the CR module. We infer this from numerical convergence in our CR acceleration rates, as well as the close conformance to analytic expectations. Nonetheless, we have tested this explicitly by checking energy absorption for the two-fluid case when $\kappa=0$ (bottom panel of Figure \ref{fig:fCR}); in this case $\dot{E}_{\rm gas}/\tilde{\epsilon} \rightarrow 0.8$ when $P_{CR}/P_{g} \sim 1$, i.e. gas heating is once again large. 

If Ptuskin damping does not allow gas motions to cascade the $\sim 2$ decades to grid scale in our simulations to enable dissipation, this strongly suggests that real turbulence should not be able to cascade down the many more decades to e.g. the gyroscale of CRs, where fast modes are frequently invoked to scatter CRs with $E \gtrapprox 300$ GeV. Of course, it is still imperative to test these ideas in much higher resolution simulations, preferably with a spectral code that can better resolve an MHD Kraichnan cascade. 

\begin{figure*}
\centering
\includegraphics[width=0.99\textwidth]{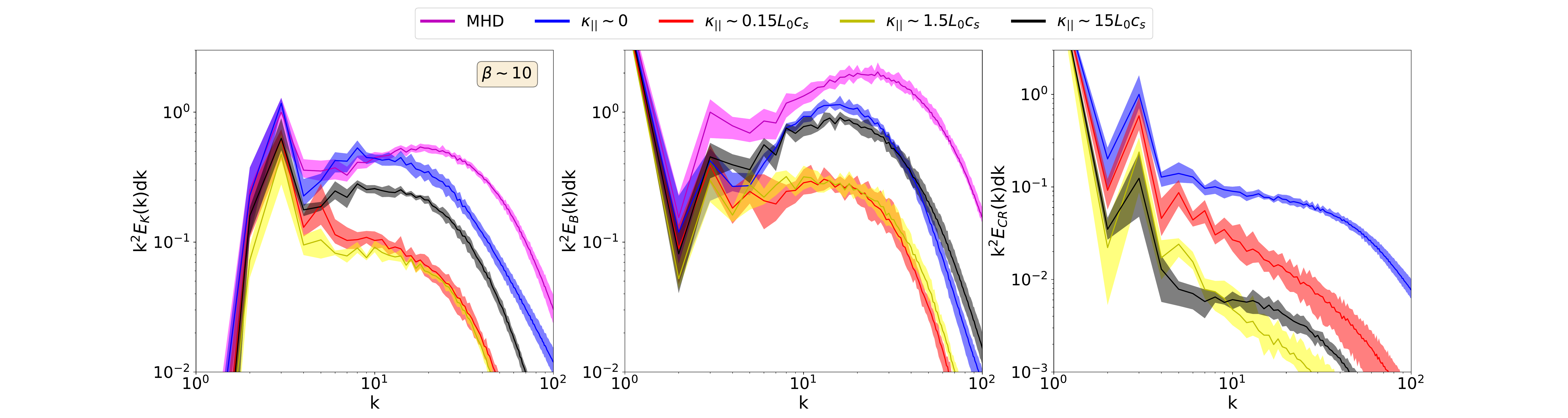}
\caption{Kinetic energy spectra (left), magnetic energy spectra (middle), and CR energy spectra (right) multiplied by $k^{2}$ for $512^{3}$, $\beta \sim 10$ simulations with varying $\kappa$. Note that the left panel is the same as in Figure \ref{fig:KE_spectra_varyKappa} and that the y-axis of the right panel extends down to $10^{-3}$ rather than $10^{-2}$ for the other panels. Overall, magnetic energy spectra follow the same trend as kinetic energy spectra, showing damped small-scale power when the diffusivity is near the sweet-spot $\kappa \sim 0.15 L_{0} c_{s}$. CR energy spectra instead show an approximately monotonic decrease in small-scale power with increasing diffusivity, as 
strong diffusion damps small-scale CR perturbations.} 
\label{fig:magnetic_energy_spectra}
\end{figure*}

While we've focused on the kinetic energy spectra so far, we have yet to show that the magnetic energy spectra, which is most important for CR scattering, shows the same damping trends. Figure \ref{fig:magnetic_energy_spectra} shows the kinetic energy spectra (left panel), magnetic energy spectra (middle), and CR energy spectra (right) for our set of $512^{3}$, $\beta \sim 10$ simulations with varying CR diffusivities. The kinetic energy spectra are identical to that in Figure \ref{fig:KE_spectra_varyKappa}, and they show considerable damping when $\kappa \sim 0.15 L_{0}c_{s}$, i.e. at the sweet-spot diffusivity where damping is most efficient. Similarly, for that same simulation, the magnetic energy spectrum is clearly damped, but as $\kappa$ varies off the sweet-spot, more small-scale power remains. The CR energy spectra are quite different: the amplitude of small-scale CR pressure fluctuations monotonically decreases with increasing $\kappa$, because strong diffusion damps small-scale CR perturbations.

\subsection{Streaming vs Diffusion}
\label{sec:streaming}

\begin{figure*}
\centering
\includegraphics[width=0.95\textwidth]{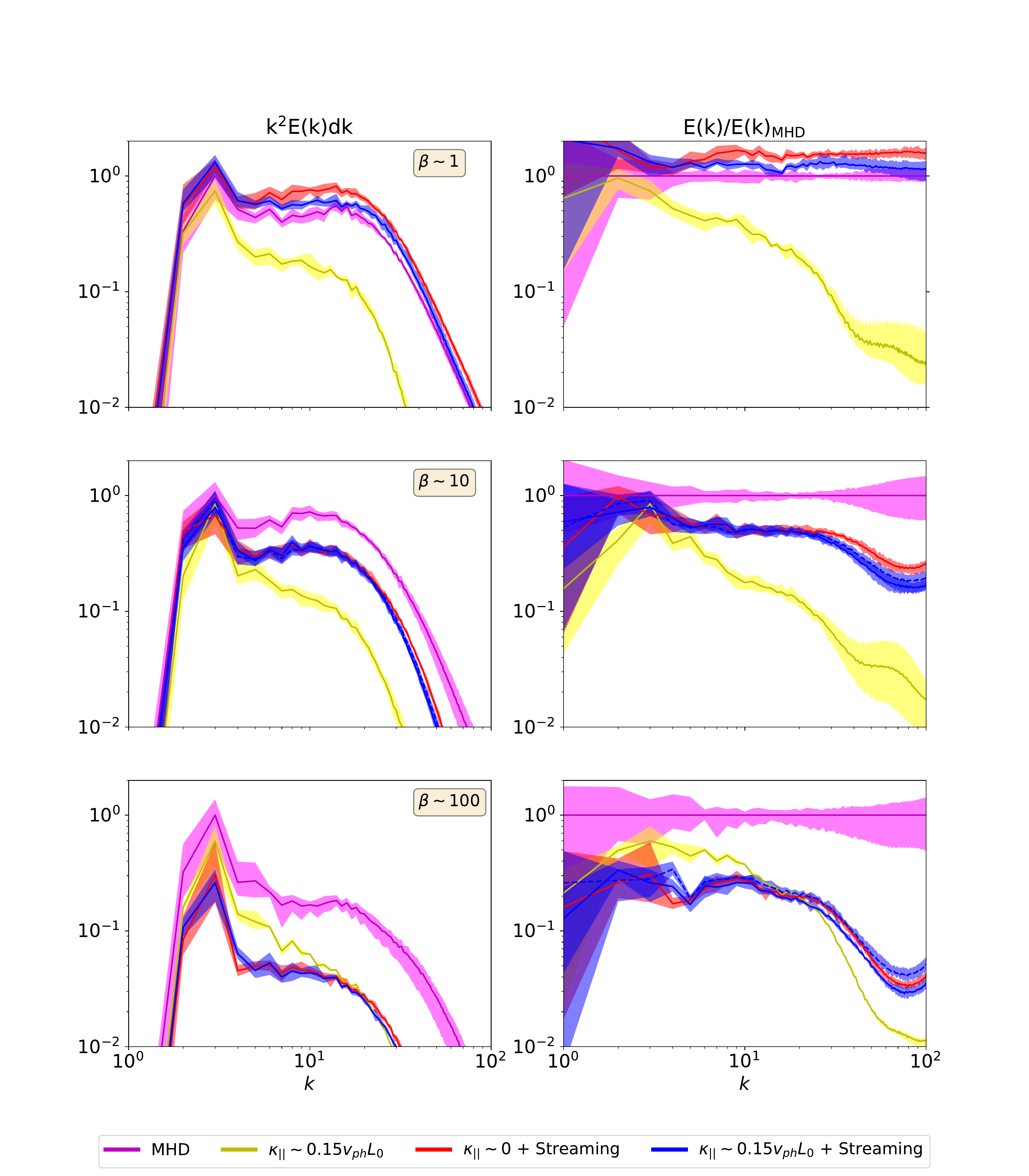}
\caption{Kinetic energy spectra for $\mathcal{M}_{\rm s} \sim 0.15$, $256^{3}$ simulations each with $P_{CR}/P_g \sim 1$ but varying CR transport and varying the initial plasma $\beta$ from 1 (top) to 10 (middle) to 100 (bottom). The magenta-colored lines show the resulting MHD (no CR) spectra as a reference. The left column shows $k^{2}E(k)dk$ normalized by the MHD value at $k=3$, while, to more clearly show the changes in spectral shape, the right column shows the ratio of each spectrum to the MHD spectrum. For diffusion only, efficient reacceleration damps the kinetic energy spectrum, resulting in less power at small scales compared to the MHD case. However, with streaming included, both reacceleration rates and field-aligned CR pressure gradients depend on $\beta$. At low $\beta$ (low Alfv\'{e}n Mach number $\mathcal{M}_{\rm A} = v/v_{A}$), streaming negates reacceleration, and the kinetic energy spectra revert to the MHD case. For larger $\beta$, however, reacceleration becomes somewhat more efficient, causing damping, and a more significant fraction of turbulent energy is channeled through CRs and lost via streaming energy transfer. This latter effect, most clearly evident in the streaming only simulations (red curves), decreases the \emph{overall} kinetic energy in the system but doesn't appear to induce cut-offs like the diffusion-only runs.} 
\label{fig:KE_spectra_streaming}
\end{figure*}

In the pure diffusion limit, $\Gamma(k)$ is well known, and as we've shown analytically and numerically, the resulting CR drag damps turbulence at large scales, changing kinetic energy spectral slopes and even introducing cut-offs. The functional form for $\Gamma(k)$ is more uncertain
when streaming transport is introduced. Since we found in Paper I that streaming stunts reacceleration rates due to fundamental changes to CR-turbulent interactions, it's tempting to append the plasma $\beta$-dependent correction factors from Paper I to $\Gamma(k)$. If this were true, weak CR reacceleration should imply very weak changes to the kinetic energy spectrum; however, we've run a number of simulations with CR streaming, including some with \emph{no diffusive transport} where reacceleration is absolutely negligible, that clearly modify the kinetic energy spectra. We present some simple scalings which match our simulations, but defer a detailed study to future work.

All simulations in this section start with $P_{CR} \sim P_{g}$ and assume an isothermal equation of state. Figure \ref{fig:KE_spectra_streaming} shows the kinetic energy spectra for $256^{3}$ simulations of varying $\beta \sim $ 1, 10, 100, each with different CR transport models but the same turbulent driving rate, which for simulations without CRs (MHD only) give a sonic Mach number $\mathcal{M}_{\rm s} \sim 0.15$. A partial version of Figure \ref{fig:KE_spectra_streaming}, using a $512^{3}$ domain, is included in the Appendix and shows similar behavior. The left column shows each spectrum multiplied by $k^{2}$, normalized to the peak value of the MHD spectrum at k=3. The right column, in order to more clearly show differences in the spectral shape and overall kinetic energy, shows each spectrum divided by the MHD spectrum. Note the similarity of the pure streaming power spectra to the streaming + diffusion power spectra; in this parameter range, streaming dominates over diffusion. We seek to answer two main questions about these results:

\begin{center}
\emph{How does streaming vs diffusive transport affect the overall kinetic energy in the gas?} 
\end{center}

\begin{figure}
\centering
\includegraphics[width=0.48\textwidth]{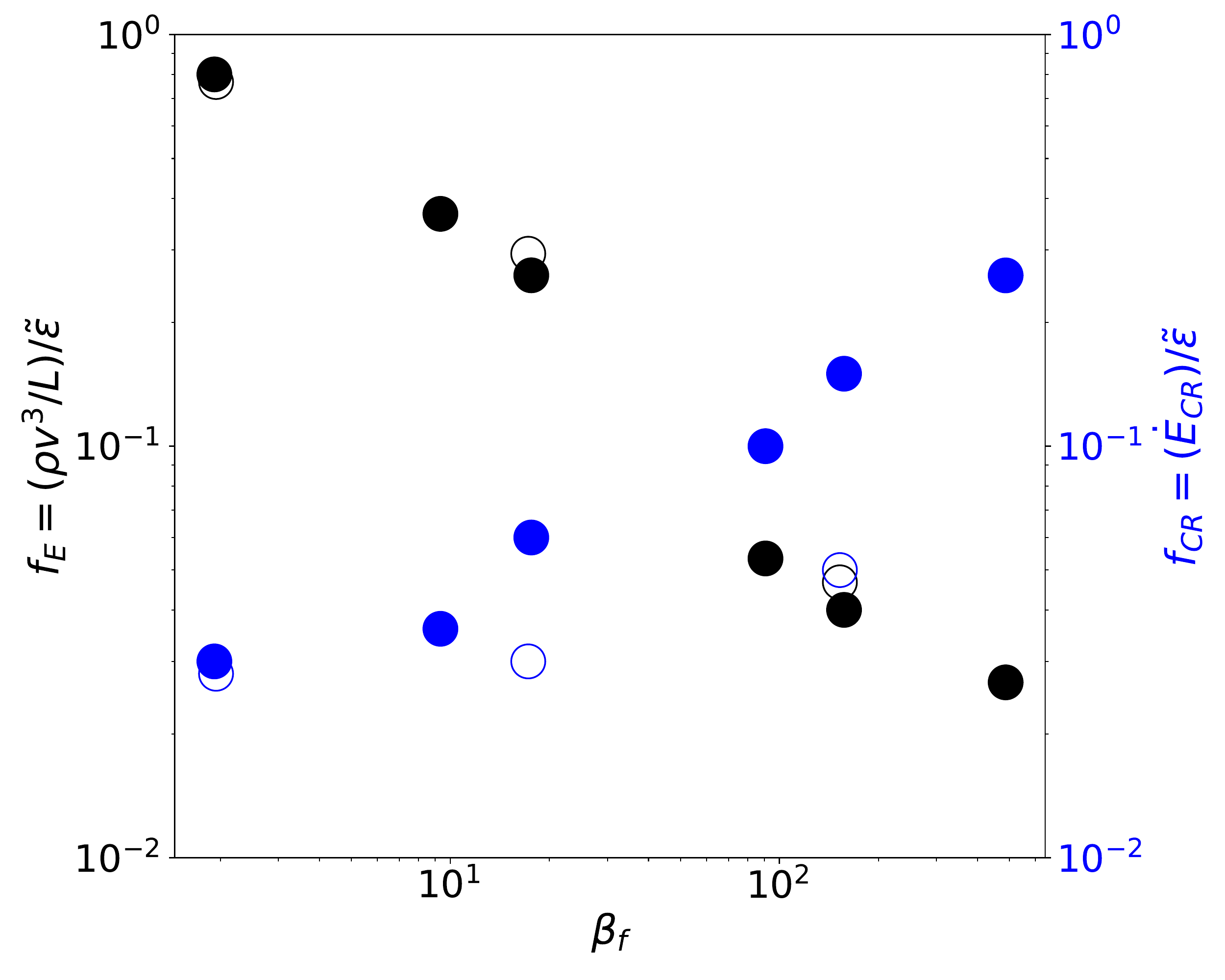}
\includegraphics[width=0.48\textwidth]{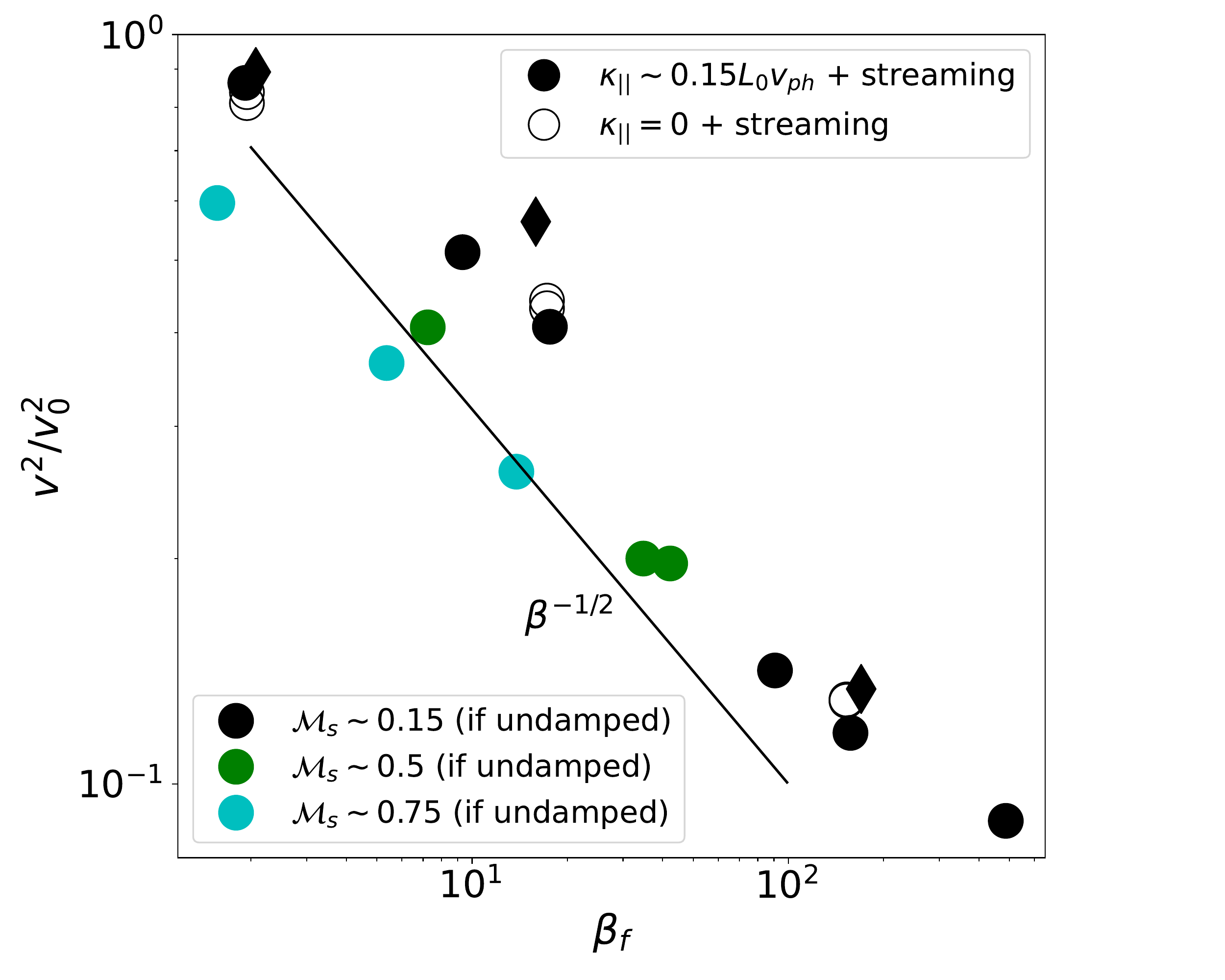}
\caption{Isothermal, $256^{3}$ (circle symbols) and $512^{3}$ (diamond symbols, bottom panel only) simulations with streaming and $P_{CR} \sim P_{g}$. The MHD (undamped) version of these simulations give $\mathcal{M}_{\rm s} = v_{0}/c_{s} \sim $ 0.15 (black), 0.5 (green), and 0.75 (cyan). The top panel shows only the $\mathcal{M}_{\rm s} \sim 0.15$ points and shows the partitioning of turbulent forcing that ends up in CRs ($f_{CR} = \dot{E}_{CR}/\tilde{\epsilon}$; green points and right y-axis), as well as $f_{E} = (\rho v^{3}/L)/\tilde{\epsilon}$ (black points and left y-axis) vs the steady-state plasma beta $\beta_{f}$. While for diffusion there was a clear correlation between $f_{CR}$ and $f_{E}$, now, $f_{CR}$ is small, and $f_{E}$ correlates inversely with $\beta$, at least in this sub-Alfv\'{e}nic regime studied. The bottom panel shows the turbulent kinetic energy relative to the undamped case, where $\rho v_{0}^{3}/L \sim \tilde{\epsilon}$. With CRs, even with streaming \emph{only} transport where there is no reacceleration, $v^{2}/v_{0}^{2} \propto \beta^{-1/2}$, at least roughly, in this sub-Alfv\'{e}nic or ``fast transport" regime. There is also a weak trend towards larger overall damping with increasing driving rate (larger $\mathcal{M}_s$ has smaller $v^2/v_0^2$), at fixed $\beta$.} 
\label{fig:Ma_vs_fE}
\end{figure}

The top panel of Figure \ref{fig:Ma_vs_fE} quantifies the partitioning of turbulent forcing that ends up in CRs ($f_{CR} = \dot{E}_{CR}/\tilde{\epsilon}$), as well as $f_{E} = (\rho v^{3}/L)/\tilde{\epsilon}$ vs the steady-state plasma beta, $\beta_{f}$. Filled circles denote simulations with streaming and diffusion, while empty circles have just streaming. The streaming plus diffusion results quantify what we see by eye in the kinetic energy spectra: increasing $\beta$ leads to smaller turbulent velocities; in each case, CRs take only a very small amount of the total energy forcing, with most energy input instead removed  from the system by streaming energy loss.

The bottom panel of Figure \ref{fig:Ma_vs_fE} shows the same simulations but with the y-axis showing the damped kinetic energy vs the undamped case. Overplotted is a line showing a $\beta^{-1/2}$ scaling, which appears to fit the data quite well. At face value, it is counterintuitive that in the streaming dominated case, CR damping is {\it stronger} at higher $\beta$, i.e. when $v_A$ is smaller. To interpret this, it's important to note that Alfv\'{e}n Mach numbers for each run saturate at $\mathcal{M}_{\rm A} < 1$, meaning that Alfv\'{e}n crossing times are faster than eddy turnover times; hence, streaming transport is relatively fast. Fast streaming transport leads to small field-aligned CR pressure gradients / large field-aligned CR scale lengths $l_{CR} = P_{CR}/(\hat{b} \cdot \nabla P_{CR})$. Compared to CRs with slow diffusive transport, streaming CRs have comparatively small pressure gradients and absorb less energy (via the $v \cdot \nabla P_{CR}$ term) in sub-Alfv\'{e}nic flows. This may partially explain the behavior seen in Figure \ref{fig:KE_spectra_streaming}, where, for instance, $\beta \sim 1$ leads to fast streaming transport, hence small CR pressure gradients, and little to no change in the kinetic energy spectrum.

At the same time, it is important to realize that CR transport timescales are {\it not} simply $\sim L/v_{\rm A}$, since CR pressure gradients and magnetic fields are often misaligned. Thus, for instance, CR heating rates (which naively scale as $\sim v_A/L$) somewhat counter-intuitively {\it decrease} as magnetic field strengths and hence $v_{\rm A}$ increase. This is because increased magnetic tension in sub-Alfvenic turbulence results in poorer alignment between magnetic fields and CR pressure gradients, reducing $v_{\rm A} \cdot \nabla P_c$ (see Figure 4 in Paper I). This qualitatively fits with the bottom panel of Figure \ref{fig:Ma_vs_fE}, assuming collisionless energy loss drives the damping.

While we do not have a rigorous argument for the $v^2/v_0^2 \propto \beta^{-1/2}$ scaling, which we present as an outcome of our simulations, we can give the following heuristic argument: $v^2 \propto \epsilon t_{\rm damp}$ (from equation \ref{eqn:veldrop}), where naively $t_{\rm damp} \propto L/v_A$. However, since CR heating rates (and hence turbulent damping rates) scale as $v_{\rm A} \cdot \nabla P_c$, we know that $t_{\rm damp}$ also depends on $P_c,P_{\rm B}$, where in sub-Alfvenic turbulence $P_B$ controls the relative alignment between $v_A$ and $\nabla P_c$ via magnetic tension. From dimensional analysis, we must have $t_{\rm damp} \sim L/v_A (P_B/P_c)^{\alpha}$, where $\alpha=1$ since $t_{\rm damp} \propto t_{\rm heat} \propto P_c^{-1}$. If so, $v^2 \propto t_{\rm damp} \propto P_B/v_{\rm A} \propto v_A \propto \beta^{-1/2}$. Future work will have to test more carefully the scalings in the ansatz $t_{\rm damp} \propto L/v_{\rm A} (P_{\rm B}/P_c) \sim v_A L/c_{\rm c}^2$ for the streaming dominated case, which closely resembles the expression $t_{\rm damp} \sim v_{\rm ph} L/c_c^2$ in the sweet spot for the diffusion dominated case. What is striking in our simulations is that CR `drag' in the streaming dominated case consistently seems to render undamped super-Alfvenic turbulence sub-Alfvenic, even though the rise in magnetic energy density (and hence rise in $v_A$) is very mild; most of the change in $M_{\rm A}$ is due to reduced gas velocities.

\begin{center}
\emph{Does streaming change the shape of kinetic energy spectra, as diffusion does?}
\end{center}
Streaming CRs, which don't themselves take an appreciable amount of turbulent energy input, still nonetheless sap kinetic energy from the system. \emph{How} the kinetic energy spectra change, however, is fundamentally different between streaming and diffusive transport. Changing $\beta$ (changing $\mathcal{M}_{\rm A}$) in streaming-dominated simulations effectively changes the ratio of transport time to eddy turnover time. To glean further insight, it's interesting to compare to simulations with purely diffusive transport but varying diffusion coefficients. 

Figure \ref{fig:KE_spectra_varyKappa} shows kinetic energy spectra for simulations on a $512^{3}$ grid, each with initial $\beta \sim 10$ but diffusion coefficients varying between $\kappa_{||} \sim (0.15-15) v_{ph} L_{0}$. Our fiducial case of $\kappa_{||} \sim 0.15 v_{ph} L_0$ shows that damping, in the slow diffusion regime, exerts meaningful drag on an entire hierarchy of scales, beginning at the outer scale; in other words, damping and cascade rates are competitive over a large range of k. Moving to the fast diffusion regime ($\kappa_{||} \sim 15 v_{ph} L_0$), this is clearly not the case: the diffusion length scale is larger than the outer eddy scale, and the damping rate is only competitive with the cascade time at large scales, leaving the cascade to operate uninterrupted after CRs have reduced the outer-scale kinetic energy. 

Following similar logic, we infer that, for streaming-dominated transport in sub-Alfv\'{e}nic turbulence, $\Gamma(k)$ must be weighted heavily towards small k, causing an initial reduction in outer-scale kinetic energy but an unimpeded cascade at larger k. Thus, we see that the power spectrum when streaming is included has the same shape over the effective inertial range of the simulations $k \lsim 30$, albeit with a lower normalization (in the $\beta \sim 10,100$ cases, when damping is effective). In the dissipation range, $k \gsim 30$, there is additional steepening compared to the MHD case, though whether this is numerical or physical is as yet unclear.

\subsection{Compressive vs Solenoidal Components}
\label{sec:comp_vs_sol}

While we intend to follow this manuscript with a larger simulation suite and more detailed analysis of CR-modified turbulence, we include a preliminary analysis here of compressive vs solenoidal motions to display some characteristics we anticipate from an expanded simulation suite. Our arguments so far have focused on CR damping of compressive fluctuations, but our kinetic energy spectra contain both compressive and solenoidal motions despite being seeded with purely compressive forcing. In hydrodynamic turbulence, compressive motions completely dominate in subsonic turbulence driven with purely compressive forcing \citep{Federrath2010}, but in MHD turbulence, magnetic fields affect this balance. Namely, for the sub-Alfvenic, $\beta \sim 10$, $M_{s} \sim 0.15$ simulations we've focused on, we expect from previous work \citep{lim2020} that the combination of compressive fluctuations and magnetic tension will generate solenoidal power, even at a level comparable to the compressive power. This holds true in our simulations.

\begin{figure*}
\centering
\includegraphics[width=0.98\textwidth]{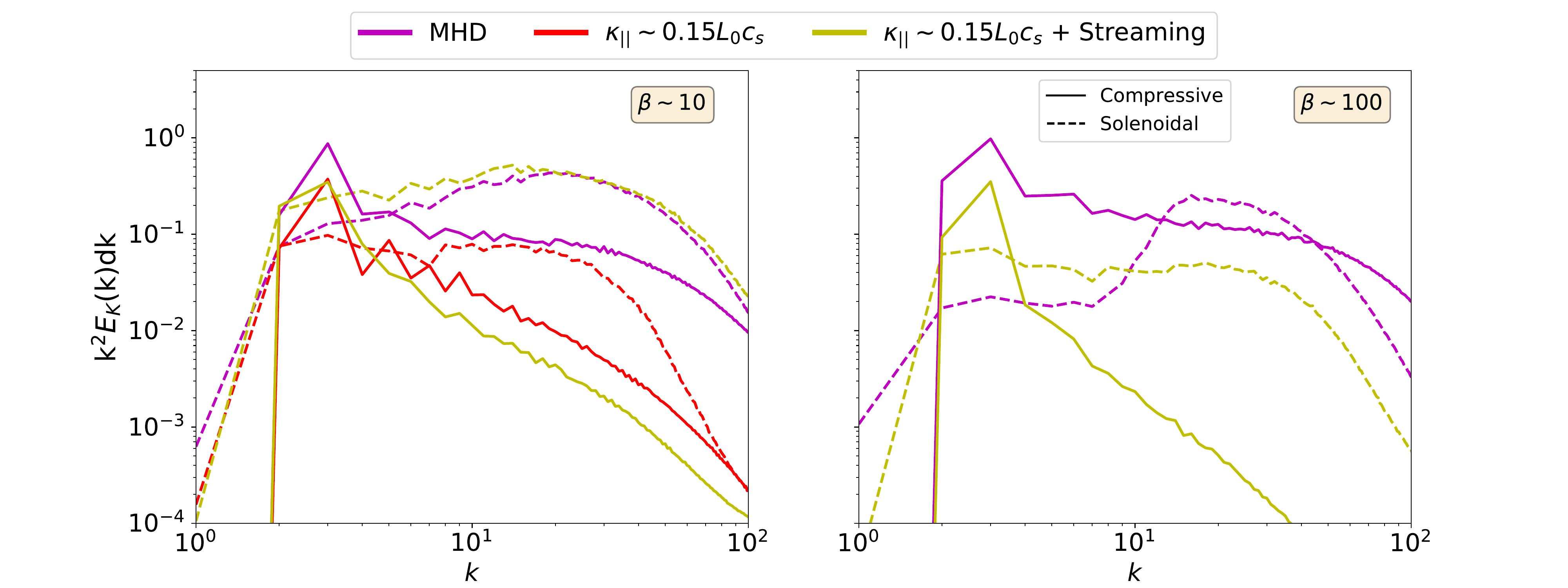}
\includegraphics[width=0.98\textwidth]{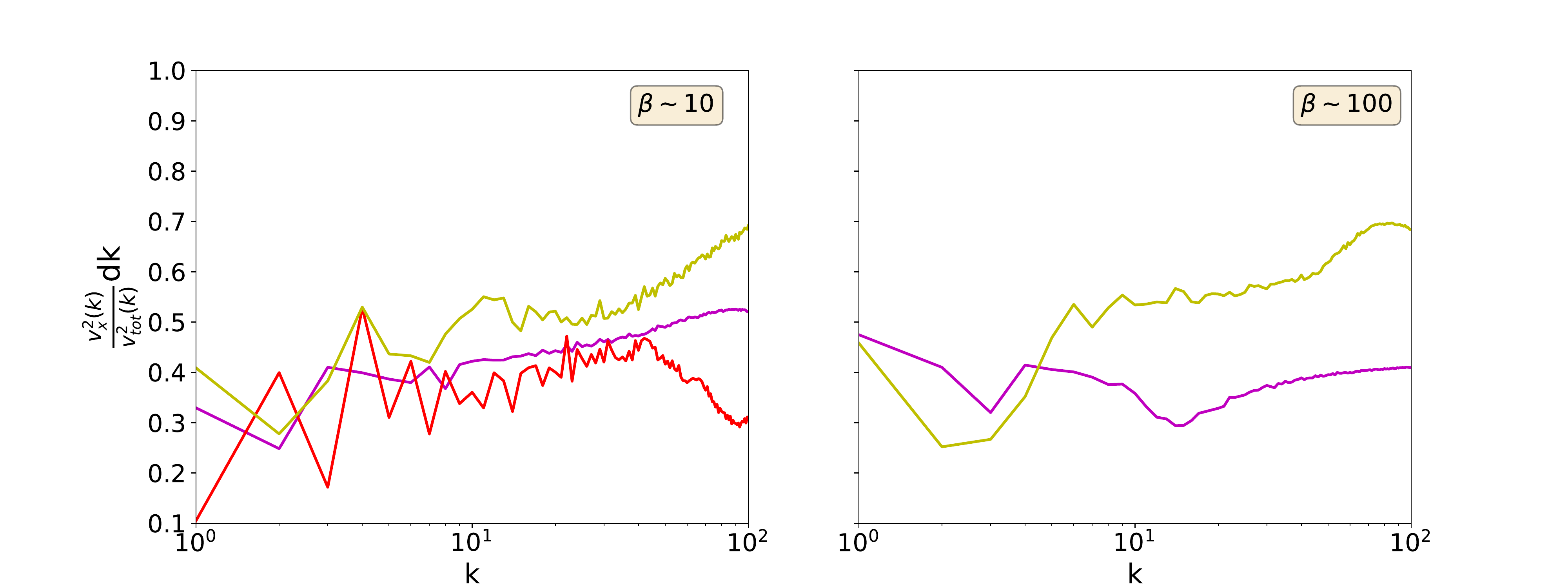}
\caption{\emph{Top row:} Kinetic energy spectra, multiplied by $k^{2}$ and normalized to the $k = 3$ MHD value, decomposed into compressive (solid lines) and solenoidal (dashed lines) components. Each simulation was run on a $512^{3}$ grid. The left panel shows $\beta \sim 10$ simulations, where solenoidal power is a significant fraction of the total power and dominates at small scales, leading to a shallower than $k^{-2}$ spectrum. The right panel shows $\beta \sim 100$ simulations, where compressive modes are dominant at almost all scales in the MHD case. CRs considerably damp the compressive fluctuations, though, which in turn decreases the power in solenoidal motions that are generated by a combination of compressions and magnetic tension. In both the $\beta \sim 10$ and $\beta \sim 100$ cases with streaming, compressive damping leads to an increased ratio of solenoidal to compressive power. However, this ``divergence cleaning" is not pronounced in the pure diffusion run, where solenoidal and compressive power decrease by about the same amount. \emph{Bottom row:} Power in velocity fluctuations $v_{x}$ along the initial mean magnetic field divided by the power in total velocity fluctuations. This quick measure of anisotropy roughly tracks the compressive vs solenoidal motions seen above: with CR streaming present, CR damping leads to more anisotropy (higher fraction of $v_{x}$ power), consistent with a larger fraction of anisotropic solenoidal modes rather than isotropic compressive modes. The diffusion only case, which shows a kink at high-k, is an outlier whose analysis we leave for future work.}
\label{fig:sol_vs_comp}
\end{figure*}

We use a standard Hodge-Helmholtz decomposition to separate compressive and solenoidal components as a function of scale, and we plot their power spectra, multiplied by $k^{2}$ and normalized by the corresponding power of the MHD simulation's $k = 3$ mode, for a subset of our $\beta \sim 10$ and $\beta \sim 100$ simulations with and without CRs in Figure \ref{fig:sol_vs_comp}. For the MHD simulations, the integrated fractions of solenoidal power to total power are $E_{\rm sol} / E_{\rm tot} \sim 0.42$ for $\beta \sim 10$ and $E_{\rm sol} / E_{\rm tot} \sim 0.11$ for $\beta \sim 100$. These values are in-line with those in \cite{lim2020}, with magnetic tension playing a small role in solenoidal generation at higher $\beta$. Interestingly, the solenoidal component is comparable to the compressive component or even dominates at small scales. That our kinetic energy spectra are shallower than $k^{-2}$ at large k, then, seems to be due to Alfven modes rather than fast modes, consistent with recent literature suggesting that, even with primarily compressive driving, significant turbulent energy lies instead in Alfven modes (see e.g. Figure 2 in \citealt{Makwana2020}, or \citealt{Gan2022} for a full spatio-temporal decomposition of fast, slow, and Alfven modes).

When CRs are present, compressive (and in some cases, solenoidal power) decreases. For the $\beta \sim 10$ case with CR diffusion, we measure $E_{\rm sol} / E_{\rm tot} = 0.36$, very comparable to the MHD case with $E_{\rm sol} / E_{\rm tot} = 0.42$. When streaming is present, we measure $E_{\rm sol}/E_{\rm tot} = 0.67$ for the $\beta \sim 10$ simulation (compare to $E_{\rm sol} / E_{\rm tot} = 0.42$ for MHD), and we measure $E_{\rm sol}/E_{\rm tot} = 0.35$ for the $\beta \sim 100$ simulation (compare to $E_{\rm sol}/E_{\rm tot} = 0.11$ for MHD). 

Our interpretation, barring further work that we save for a future paper, is that CRs preferentially damp compressive motions consistent with the analytic derivations of this paper, but since compressive motions combine with magnetic tension to drive solenoidal motions in sub-Alfvenic turbulence, both compressive and solenoidal components are suppressed. That CRs preferentially damp compressive rather than solenoidal motions is evidenced by our two simulations with CR streaming, which show a ``divergence cleaning" effect where the ratio of solenoidal to compressive power increases. However, this divergence cleaning is less apparent in the diffusion only run. We defer a fuller discussion of the difference between diffusion and streaming effects to later work where we drive both compressive and solenoidal modes, rather than relying on solenoidal motions generated by compressive forcing, since in this case damping of compressive motions can easily damp solenoidal power as well.

For brevity, we defer most other analyses of the cascade to future work, but we do point out one additional outcome of damping: CRs can change the anisotropy of the cascade. In the bottom row of Figure \ref{fig:sol_vs_comp}, we plot the power in velocity fluctuations along the initial magnetic field direction ($\hat{x}$) over the power in all directions. This quick diagnostic of anisotropy follows our intuition from above: In the simulations with CR streaming, as CRs damp isotropic compressive motions, and turbulence is dominated by anisotropic solenoidal motions, eddies become more elongated along the mean field direction, with the fractional power in $v_{x}$ fluctuations increasing from the MHD values, especially in $k = 5-20$ range. There is no such increase in anisotropy in the diffusion only run, consistent with the lack of divergence cleaning. Indeed, there is an apparent downturn toward isotropy at high k, albeit in a range where results may not be numerically reliable.

\section{Discussion}
\label{sec:discussion}
\subsection{Regimes of CR Modified Turbulence}

In \S\ref{sec:spectra}, we showed both analytically and numerically that CRs can have a significant impact on the power spectrum of turbulence. In particular, we showed that as the damping time decreases relative to the turbulent cascade time, the turbulent power spectra will be steepen and then cut off abruptly at small scales (for $t_{\rm cascade}/t_{\rm damp} \gsim 1$).  
These results should eventually be carefully checked by higher resolution numerical simulations. Nonetheless, we clearly already have seen in \S\ref{sect:CR-diversion} and Fig \ref{fig:fCR} a situation where CR damping of motions is stronger than the rate at which energy cascades to smaller scales, so that little energy reaches the grid scale. Our analytic estimates can guide expectations as to which environments these effects might be important. 

\paragraph{Intra-cluster medium (ICM)} In the ICM, although sonic Mach numbers are typically low ($\mathcal{M}_{s} \sim 0.1-0.3$), the absence of hadronic $\gamma$-ray emission gives an upper bound on $P_c/P_{\rm tot} << 1$ (typically less than a few percent; \citealt{ackermann14}), so that Equation \ref{eq:tdamp_min} is not satisfied there. The CR energy density is too small to appreciably affect gas motions, and it is unlikely that CR reacceleration appreciably damps the turbulent cascade.

\paragraph{Interstellar medium (ISM)} In the ISM, CR damping could be potentially important: $P_{\rm c}/P_{\rm tot} \sim \mathcal{O}(1)$ is relatively large. In the diffusion only case, the main uncertainty lies in the CR acceleration rate. The most efficient reacceleration occurs for diffusivities in the range $\kappa_{||} < v_{ph} L_{0} \sim 3 \times 10^{26} \rm cm^{2}/s$ for ISM-like parameters (see Table 2 in Paper I). Canonical values of $\kappa \sim 10^{28}-10^{29} \rm cm^{2}/s$ used in galactic propagation models are much larger, i.e. we are sufficiently far away from the `sweet spot' that acceleration and hence damping times could be long. On the other hand, if CR streaming dominates transport, then since the ISM has $\beta \sim 1$, damping is small, as we have seen.

\paragraph{Circumgalactic medium (CGM)} Finally, the galactic halo and CGM are strong candidates for significant CR damping. For the diffusion only case, these regions occupy a sweet-spot where $\kappa \sim v_{\rm ph} L_0$, $\mathcal{M}_{\rm ph} < 1$, and if, as suggested by simulations of Milky Way mass galaxies (e.g. \citealt{butskyWinds2018, JiCRHalos2020}), $P_{\rm c}/P_{\rm g} \gsim 1$, then $P_{\rm c}/P_{\rm tot}$ is order unity. For these conditions, Equation \ref{eq:tdamp_min} is satisfied, so that $t_{\rm damp} \sim t_{\rm inject}$. Thus, for instance, from Equation \ref{eq:ratio_t_cascade_damp}, $t_{\rm cascade}/t_{\rm damp} \sim \mathcal{M}_{\rm ph}^{-1} \sim 2$ for a compressive Kraichnan cascade with $\mathcal{M}_{\rm ph} \sim 0.5$: the compressive cascade will be steepened beyond the critical threshold of $E(k) \propto k^{-2}$ and abruptly cut off, so there is no small scale turbulence. We see hints of this in Figure \ref{fig:CR_Turb_Modify} for the $\mathcal{M}_{\rm s} \sim 0.5$ case, but given our limited dynamic range, spectral changes are more obvious for $\mathcal{M}_{\rm s} \sim 0.15$, when $t_{\rm NL}/t_{\rm damp}$ is even larger. 

Once streaming is included, we have also seen that there can be considerable damping in the $\beta \sim 10-100$ cases, with a weak trend towards larger damping with increasing driving rate at fixed $\beta$ (see how $v^2/v_0^2$ is smaller for larger $\mathcal{M}_s$), potentially because CRs are more efficiently trapped in turbulent eddies as the Alfven Mach number approaches unity. This differs from the diffusion only case, where stronger turbulence implies smaller $t_{\rm NL}/t_{\rm damp}$ and weaker damping. For a fixed driving rate that produces transonic MHD turbulence in a $\beta \sim 10$ environment (reasonable CGM parameters), adding streaming-dominated CRs up to equipartition $P_{CR} \sim P_{g}$ damps turbulent kinetic energy by a factor of $\sim 5$ or greater (bottom panel of Figure \ref{fig:Ma_vs_fE}). 

For each of these regimes, there is also a question of the turbulent driving scale relative to the CR mean free path, i.e. whether our fluid assumption of CR transport is valid. For both self-confinement and extrinsic turbulence models of CR scattering, the typical mean free path for a GeV CR in the ISM is $\sim 1$ pc, which is not too far below the typical driving scale of turbulence ($\sim 100$ pc). However, if self-confinement is stronger, then the mean free path is shorter, and the separation between driving scale and mean free path is larger. Similarly, in the CGM and ICM, the driving scale is much larger, so this scale separation is not an issue.

Finally, given the possibilities for CR-modified turbulence in ISM and CGM environments described above, how do these results compare to observations of electron density fluctuations measured through interstellar scintillation \citep{BigPowerLaw}, which show density fluctuations on a wide range of scales, i.e. the ``Big Power Law" in the sky? We believe our results are consistent with these observations for two reasons: (i) The observed spectrum $\propto k^{-5/3}$ is consistent with Kolmogorov turbulence and is, therefore, unlikely to be generated by a purely compressive fast mode cascade. From our findings, CRs preferentially damp compressive fluctuations (\S\ref{sec:comp_vs_sol}), allowing solenoidal motions to extend over a wide range of scales consistent with the ``Big Power Law". (ii) In any case, in the $\beta \sim 1$ ISM, if streaming is dominant, CRs do not significantly modify the power spectrum (Figure \ref{fig:KE_spectra_streaming}). Signatures of small scale damping are more likely to be seen in the $\beta > 1$ CGM, if the CR energy density is significant (as is suggested by simulations; \citealt{JiCRHalos2020}).

\subsection{Implications of CR-Modified Turbulence}
The implications of such CR-modified spectra are possibly quite intriguing. For instance, a CR-induced cut-off could significantly affect the spatial scale of thermal instability, since there are no small scale compressive motions, unless there is direct driving at those scales. Also, since Ptuskin damping only affects compressive motions, not solenoidal motions, Ptuskin damping can potentially make turbulence less Burgers-like and more Kolmogorov-like. It would be interesting to explore this `divergence-cleaning' effect in simulations with a mixture of driving modes.

Perhaps the most interesting consequence of CR damping of turbulence is its implication for scattering of high-energy CRs by fast modes in an extrinsically driven turbulent cascade. This is frequently invoked to explain the scattering of CRs with $E \gsim 300$GeV \citep{Yan2004}, since self-confinement is too weak to explain observed isotropy and confinement times. However, the resonant scattering invoked (transit time damping) requires the turbulence to cascade many orders of magnitude, to the $\sim 300$AU gyroscale of such CRs. Fig \ref{fig:dynamic} shows that for $t_{\rm casc}/t_{\rm damp} = 1$, a Kraichnan ($E(k) \propto k^{-3/2}$) fast mode spectrum will steepen to a Burgers ($E(k) \propto k^{-2}$) spectrum, which already has too little small scale power to efficiently scatter CRs via transit time damping \citep{miniati15,Pinzke2017}, and even higher values of $t_{\rm casc}/t_{\rm damp} \sim \mathcal{M}_{\rm ph}^{-1}$ will completely eliminate turbulence at small scales. While this needs further study, low-energy CRs, by damping turbulent fluctuations at large scales, could divert turbulent energy that would otherwise scatter high-energy CRs. This potentially adds to the long list of problems with `standard' theories of CR scattering in the Milky Way which have been recently pointed out \citep{Kempski2022,Hopkins2022_propblems}. 

Regardless of whether CR drag introduces a cut-off to kinetic energy spectra, it is clear from our simulations that CRs in both diffusion-dominated and streaming-dominated transport regimes can sap a significant fraction of the turbulent forcing rate. This breaks the usual correspondence between turbulent velocity and turbulent driving rate, i.e. for hydrodynamic turbulence, $\rho v_{0}^{3}/L \sim \tilde{\epsilon}$. Now, $\rho v^{3}/L \sim f_{E} \tilde{\epsilon}$, where the new correction factor $f_{E}$ can be $\ll 1$. As derived in Equation \ref{eqn:veldrop}, $v^{2}/v_{0}^{2} \propto t_{\rm damp}/t_{\rm cascade}$, which for streaming-dominated transport in sub-Alfv\'{e}nic turbulence gives $v^{2}/v_{0}^{2} \propto \beta^{-1/2}$ (Figure \ref{fig:Ma_vs_fE}). In the CGM, where we expect damping to be most significant, \emph{turbulent velocities obtained from the observed velocity dispersion may significantly underestimate the turbulent forcing rate, i.e. $\tilde{\epsilon} \gg \rho v^{3}/L$.}

\section{Conclusions}
\label{sec:conclusions}

In this paper, we present analytical estimates and accompanying MHD+CR simulations probing CR effects on turbulence, namely the damping of turbulence by large-scale, CR-induced drag on compressive gas motions. Our main findings are as follows:

\begin{itemize}[left=4pt]
    \item Despite long CR reacceleration times, the damping time due to CR reacceleration can be very competitive with the turbulent cascade time.
    \begin{equation}
        t_{\rm damp} \sim \rho v^{2} \rm max \left( \frac{t_{\rm grow}}{P_{\rm CR}}, \frac{1}{\tilde{\epsilon}} \right) \sim \rm max \left(\mathcal{M}_{\rm c}^{2} t_{\rm grow},\frac{\rho v^{2}}{\tilde{\epsilon}} \right)
        \label{eq:tdamp_conclusions}
    \end{equation}
    where $M_{\rm c} = v/c_{\rm c}$ is the Mach number with respect to the CR sound speed $c_c \sim \sqrt{P_c/\rho}$, and $\tilde{\epsilon} = \rho v^{3}/L$ is the turbulent energy injection rate. Our key figures are Figures \ref{fig:fCR} and \ref{fig:barchart_iso}, where we confirm that CRs can divert a significant fraction of turbulent energy that would otherwise dissipate as heat at small scales. Conditions for strong damping are met under quite reasonable conditions (Equation \ref{eq:tdamp_min}); the CGM is an especially strong candidate for this damping. 
    
    \item If CR diffusion dominates transport, and if the ratio of the damping time to the cascade time is sufficiently short, small scale compressive turbulence should be exponentially suppressed (see Figure \ref{fig:dynamic}). This suppression of small scale turbulence
    has abundant implications for e.g. thermal instability, ``divergence-cleaning" of turbulence spectra (e.g. Figure \ref{fig:sol_vs_comp}), and suppression of fast modes at small scales, which have been invoked to scatter high-energy CRs (see \S \ref{sec:damping_analytics}). We see compelling signatures of damping in our simulation spectra (\S \ref{sec:damping-sims}; Figure \ref{fig:CR_Turb_Modify}), but these effects deserve future study with higher resolution simulations that capture a larger turbulent inertial range.
    
    \item The effects of streaming transport are more complex and deserve follow-up. Importantly, $t_{\rm grow}$ in Equation \ref{eq:tdamp_conclusions} does {\it not} include the suppression of CR reacceleration by streaming (the $\beta$ dependent factors identified in \citealt{BustardOh2022_reacceleration}), which would substantially increase damping times. Instead, from Figure \ref{fig:barchart_iso}, diversion of turbulent energy through CRs remains strong even in the presence of CR streaming, for our simulations where $M_{\rm A} \lsim 1$. Instead of introducing spectral cut-offs, streaming uniformly decreases the normalization of the turbulent power spectrum, but not its shape, with the turbulent kinetic energy scaling as $v^{2} \propto v_{\rm A} \propto \beta^{-1/2}$ (Fig \ref{fig:KE_spectra_streaming}, Fig \ref{fig:Ma_vs_fE}). This is possibly because damping operates predominantly at the largest scales in the `fast transport' regime (here, the sub-Alfv\'{e}nic regime). Such large scale damping implies energetic input and turbulent heating rates (much of which gets channeled into CR collisionless heating) can be much larger than standard estimates for Kolmogorov turbulence, $\tilde{\epsilon} \gg \rho v^{3}/L$. 
    
\end{itemize}

\section*{acknowledgments}

The authors gratefully acknowledge Navin Tsung, Max Gronke, Yan-Fei Jiang, Christoph Federrath, Hui Li, and Ellen Zweibel, as well as the organizers and participants of the KITP ``Fundamentals of Gaseous Halos" workshop. We also thank our anonymous referee for an extremely detailed and perceptive report that significantly improved our paper. CB was supported by the National Science Foundation under Grant No. NSF PHY-1748958 and by the Gordon and Betty Moore Foundation through Grant No. GBMF7392. SPO was supported by NSF grant AST-1911198, and NASA grant 19-ATP19-0205.

Computations were performed on the Stampede2 and PSC-Bridges2 supercomputers under allocation TG-PHY210004 provided by the Extreme Science and Engineering Discovery Environment (XSEDE), which is supported by National Science Foundation grant number ACI-1548562 \citep{xsede}.

\software{Athena++ \citep{AthenaRef}, yt \citep{ytPaper}, Matplotlib \citep{matplotlib}, Mathematica \citep{Mathematica}}

\appendix

\section{Turbulent Properties and Damping in a Cosmic Ray Dominated Medium}
\label{appendix:turb}

CRs can influence velocity and density perturbations in a turbulent medium, but the extent depends on the relative partition of CR vs thermal energy, as well as the CR diffusivity / transport speed. \cite{commercon2019} simulate CRs in a turbulent box with purely diffusive transport and a bi-stable ISM (with radiative cooling). They found that trapped CRs modify the gas flow, change the density PDF, and provide support against thermal instability, maintaining the gas in an intermediate temperature state that is classically thermally unstable. It remains to be seen how these simulations would change when streaming is included. The perturbative heating term from CR streaming affects thermal instability \citep{Kempski2020}, and in low-$\beta$ plasmas where this heating is most significant, CR streaming can also drive acoustic waves unstable, generating a ``stair-case" cosmic ray pressure profile and additional multiphase gas \citep{Tsung2021_staircase, quataert21-streaming}. 

\begin{figure}
\centering
\includegraphics[width=0.65\textwidth]{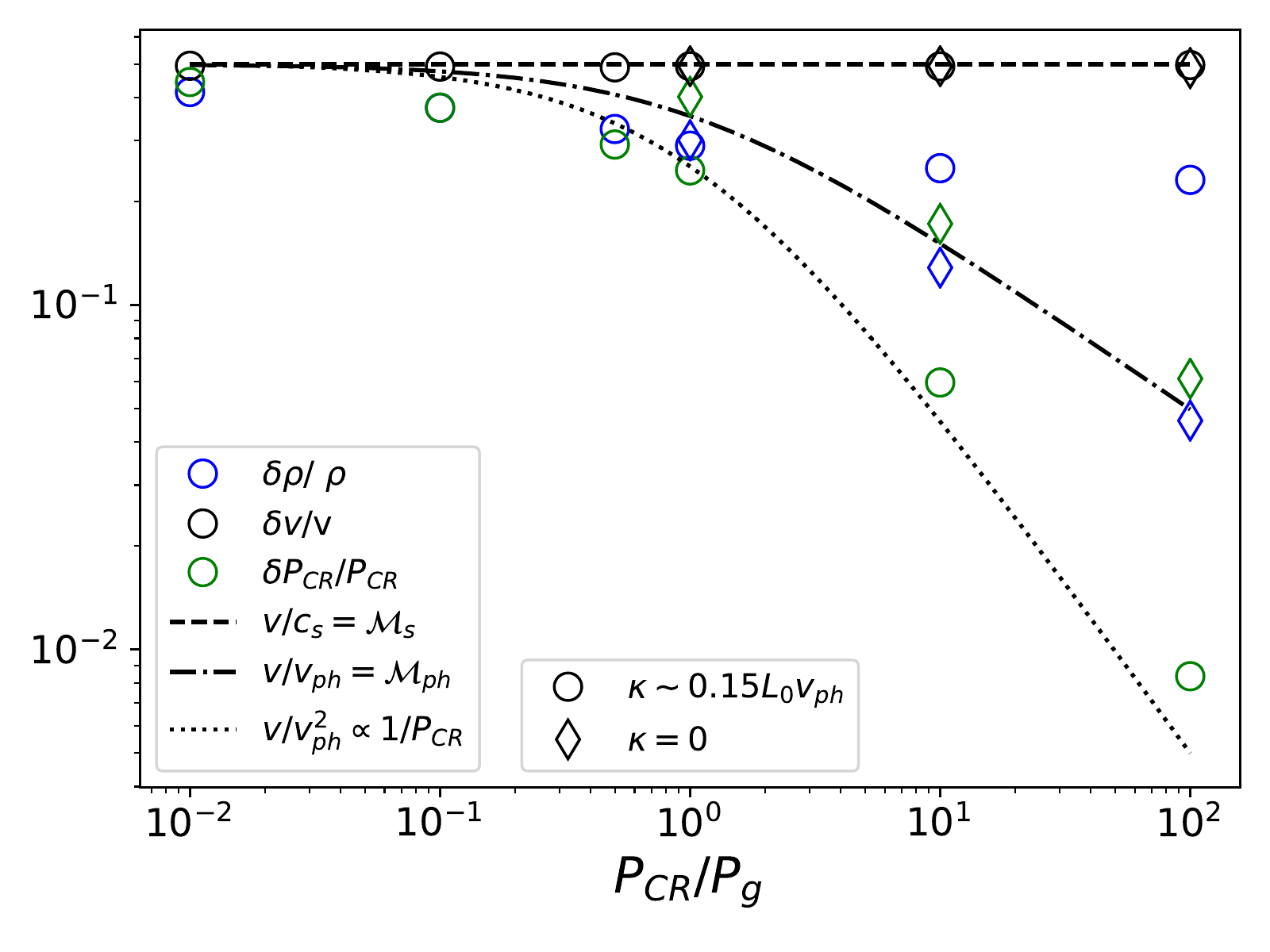}
\caption{Fluctuating density (blue symbols), velocity (black symbols), and CR pressure (green symbols) as a function of $P_{CR}/P_{g}$. Open circles denote simulations with non-zero diffusion coefficient $\kappa_{||} \sim 0.15 L_{0} v_{ph}$, and open diamonds denote purely advective CR transport ($\kappa = 0$). None of these simulations include additional streaming transport.}
\label{fig:deltarho}
\end{figure}

Our simulation setup is quite different from that of \cite{commercon2019}, most notably because we don't include radiative cooling, so we don't attempt a detailed comparison, but we do find some qualitatively similar behavior. Figure \ref{fig:deltarho} shows $\delta \rho / \rho$, $\delta v / v$, and $\delta P_{CR} / P_{CR}$ for diffusion-only simulations with varying $P_{CR}/P_{g}$ and either $\kappa = 0$ or $\kappa = 0.15 L_{0} v_{ph}$ (where $v_{ph}$ depends on $P_{CR}/P_{g}$). When CRs are dynamically unimportant ($P_{CR}/P_{g} \ll 1$), we recover the MHD expectation that $\delta \rho / \rho \sim \delta v / v \sim \mathcal{M}_{s} = 0.5$. Deviations from this relation start when $P_{CR}/P_{g} \gtrapprox 1$. Interestingly, we find that $\delta \rho/\rho$ is {\it independent} of $P_c/P_g$, while $\delta P_c/P_c \propto 1/P_c$ (i.e., $\delta P_c \sim$const is independent of $P_c/P_g$). At the same time, we find that the velocity divergence $\nabla \cdot v  \propto P_c^{-1/2}$ (not shown), i.e. it {\it does} depend on $P_c$. This might appear puzzling, since one expects density fluctuations and velocity divergence to be directly related, yet the former is independent of $P_c$, while the latter shows dependence. 

A key to understanding these results is to realize that the `sweet spot' $\kappa \sim L v_{\rm ph}$ is really still in the `fast diffusion regime'. The ratio $t_{\rm diffuse}/t_{\rm sc} \sim l v_{ph}/\kappa$ is only unity at the outer scale $l \sim L$; at smaller scales, $t_{\rm diffuse}/t_{\rm sc} < 1$ and diffusion dominates. In this diffusion dominated regime, CRs diffuse out of eddies before they contribute significantly to resisting compression -- i.e., they do not provide a significant restoring force (instead, they provide drag). In particular, they do not contribute to the phase velocity $v_{\rm ph}$. Thus, $\delta \rho/\rho \sim \delta P_g/P_g \sim v/c_s$, where $c_s$ is the gas sound speed, independent of $P_{CR}$. This is roughly consistent with \cite{commercon2019} (see their Figures 5 and 6), which finds a similar dependence on $\kappa$ and a clear decrease in $\delta P_{CR}/P_{CR}$ as $P_{CR}/P_{g}$ increases.

Using this information, we can better interpret the lower bound on damping time that we infer from our simulations. Importantly, as Ptuskin damping saturates ($P_{CR}/P_g \rightarrow \infty$, $f_{\rm CR} \rightarrow 1$), the {\it maximum} rms CR pressure perturbation is $\langle \Delta P_{CR} \rangle_{\rm rms}  \sim \rho v^2$. This is a strict upper bound, since the free energy to create CR pressure perturbations is derived from kinetic energy (similarly, $\Delta P_{CR}, \Delta P_g$ at a shock cannot exceed the ram pressure $\rho v^2$). In this limit, $\Delta P_{CR}/P_{CR} \sim \rho v^{2}/P_{CR} \propto 1/P_{CR}$. Finally, in the diffusion dominated limit, CR compression is balanced by diffusion, $P_{\rm CR,0} (\nabla \cdot v) \sim - \nabla \cdot (\kappa \nabla P_{\rm CR,1}) \sim \kappa \rho v^{2}/L^{2}$, which implies that 
\begin{equation}
    \nabla \cdot v \propto \frac{\kappa}{P_{\rm CR,0}},
\end{equation}
where $P_{\rm CR,0}, P_{\rm CR,1}$ refer to the unperturbed and perturbed CR pressure, respectively. Thus, in the regime where we fix the sweet-spot diffusion coefficient $\kappa \sim v_{\rm ph} L_0$ and $P_{CR}/P_g \gsim 1$ (so that $v_{\rm ph} \propto P_{CR}^{1/2}$), then $\kappa \propto P_{CR}^{1/2}$, and $\nabla \cdot v \propto P_{CR}^{-1/2}$. We have also verified in our simulations that $\nabla \cdot v \propto \kappa$ for constant $P_{CR}$, and $\nabla \cdot v \propto 1/P_{CR}$ for constant $\kappa$.

These results thus indicate that the damping time cannot become arbitrarily small. If drag forces are given by: 
\begin{equation}
    \dot{v} \sim \frac{1}{\rho} \nabla P_{\rm CR,1} \lsim \frac{v^{2}}{L}
    \label{eq:vdot_drag} 
\end{equation}
(where $P_{\rm CR,1} \lsim \rho v^{2}$), this gives a damping time $t_{\rm damp} \sim v/\dot{v} \gsim L/v \sim t_{\rm eddy}$. Thus, $\delta P_{\rm CR} \lsim \rho v^{2}$ implies that $t_{\rm damp} \gsim t_{\rm eddy}$. This is equivalent to the statement that the work done by CR forces in opposing gas motions cannot exceed the energy input rate: $v \cdot \nabla P_{\rm CR,1} \lsim \tilde{\epsilon} \sim \rho v^{3}/L$, which implies $\nabla P_{\rm CR,1} \lsim \rho v^{2}/L$, consistent with Equation \ref{eq:vdot_drag}. Thus, for Kolmogorov turbulence, we expect $t_{\rm cascade}/t_{\rm damp} \sim 1$ for maximally efficient Ptuskin damping. In the Kraichnan case, however, $t_{\rm cascade}/t_{\rm damp}$ can be greater than 1 if the Mach number, relative to the velocity of compressible fluctuations, is small. This is not due to any decrease in the damping time; instead, it is due to cascade times being lengthened when $v_{ph}$ is large.

\section{Time Convergence}

\begin{figure*}
\includegraphics[width=0.48\textwidth]{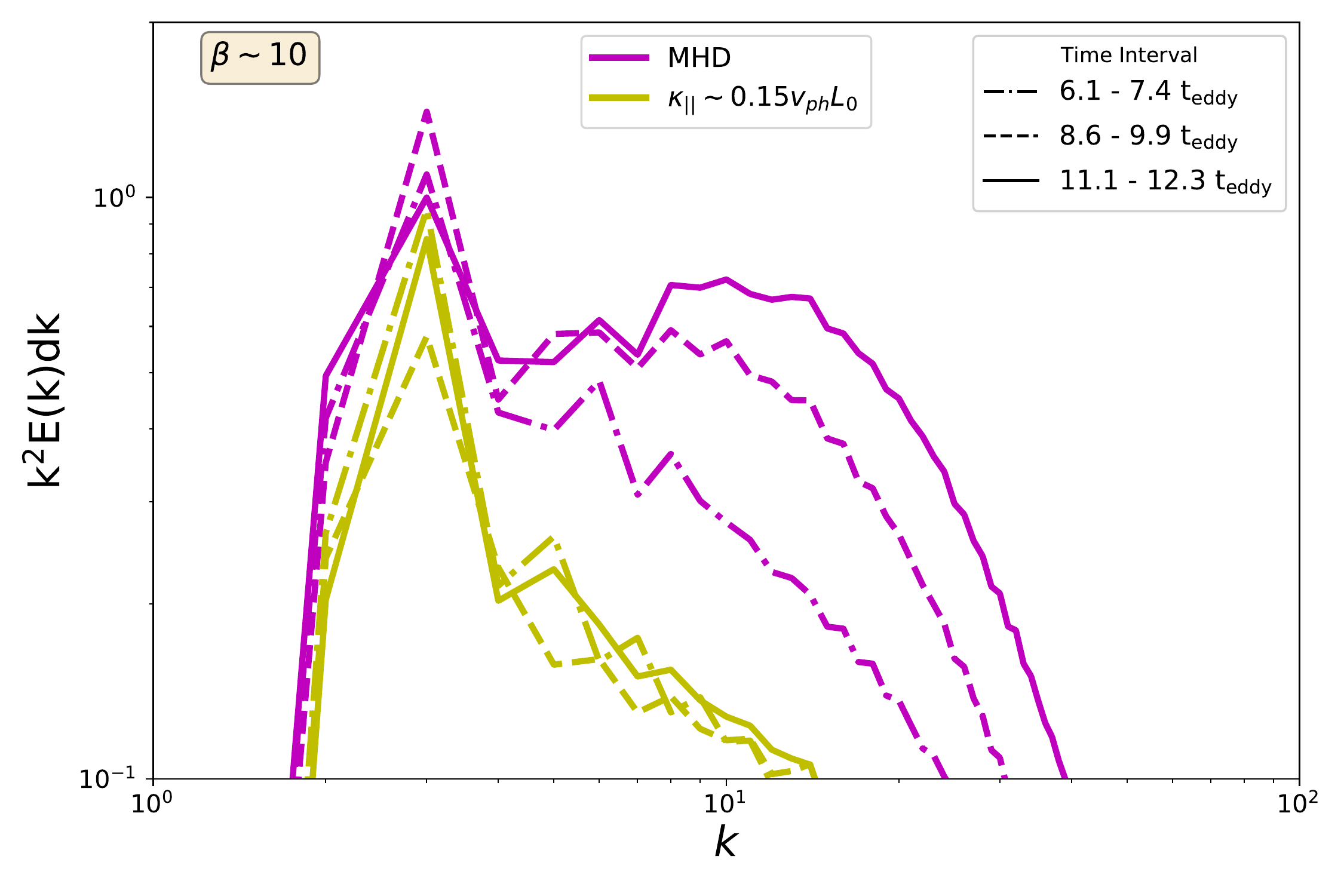}
\includegraphics[width=0.48\textwidth]{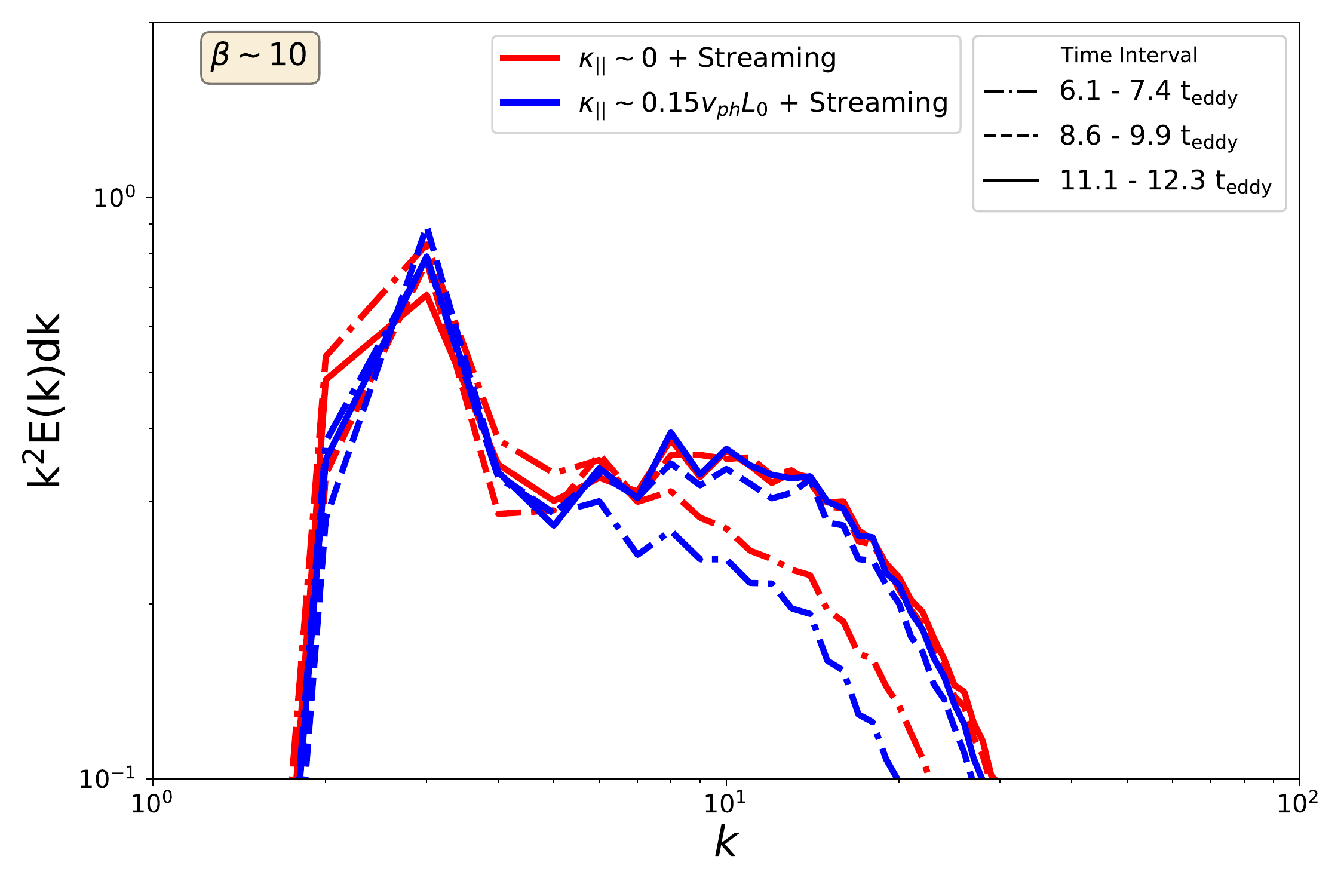}
\caption{Time convergence of select spectra, each run on a $256^{3}$ grid. The diffusion-only simulations, which show the most damping, converge very early. $512^{3}$ simulations (not shown) are similarly converged with respect to time.}
\label{fig:timesteady_spectra}
\end{figure*}

Figure \ref{fig:timesteady_spectra} shows the average kinetic energy spectra for $256^{3}$ simulations with varying CR transport model, measured at different time intervals. Most importantly, the diffusion-only simulations show converged, clearly damped spectra even at early times. Spectra for simulations with CR streaming are also well-converged but at somewhat later times. Note that these time intervals over which we pull out kinetic energy spectra are much later than the saturation of \emph{bulk} turbulent quantities (e.g. kinetic energy, magnetic energy, etc.), which occurs after only a few eddy turnover times. 

\section{Resolution Convergence}
\begin{figure}
\includegraphics[width=0.95\textwidth]{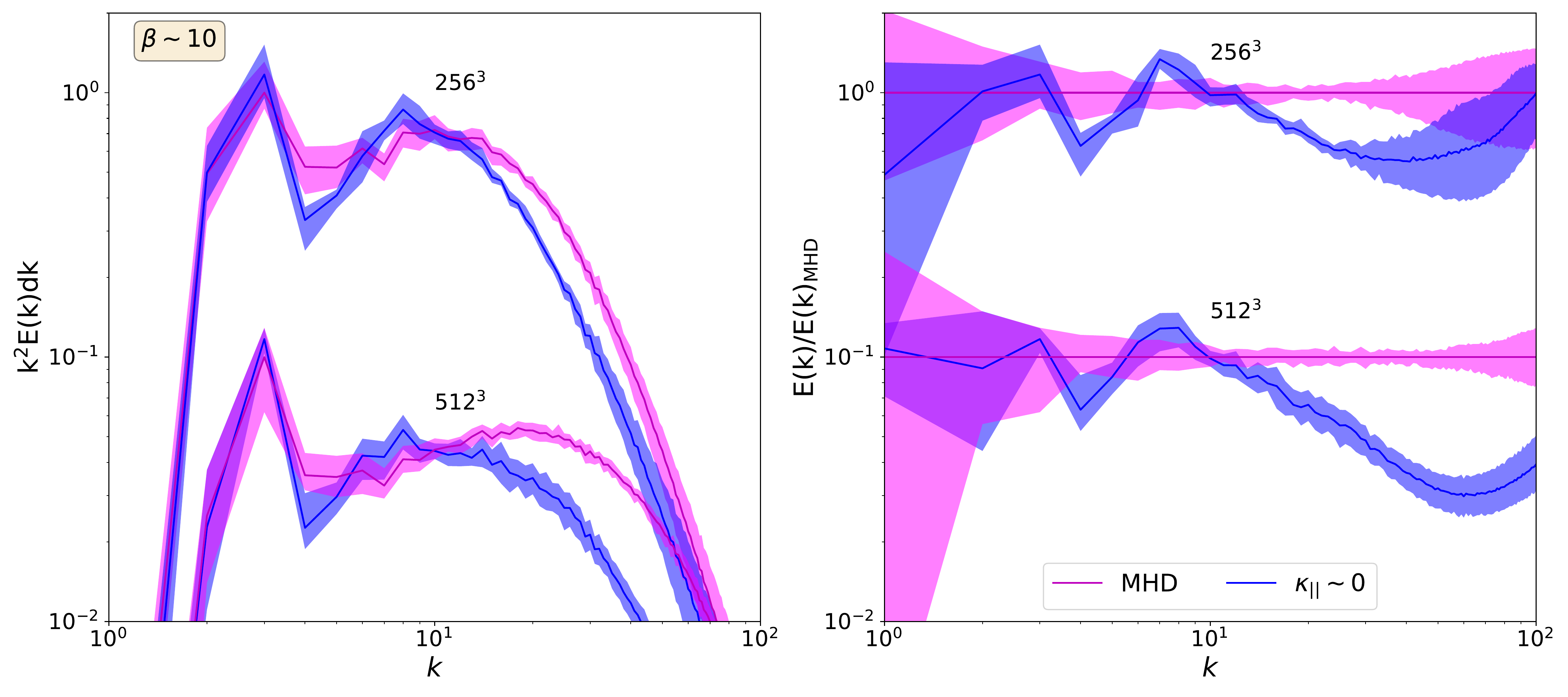}
\caption{Comparison of kinetic energy without CRs (MHD) and with CRs but no transport ($\kappa_{||} \sim 0$), simulated on grids with $256^{3}$ and $512^{3}$ cells. $512^{3}$ simulation results are divided by a factor of 10 to separate those curves from the $256^{3}$ results.}
\label{fig:varyRes_noTransport}
\end{figure}

A good test of how inherently diffusive our CR module is, and whether that accounts for some observed spectral changes, is to run simulations with no explicit CR diffusion at various resolutions. Figure \ref{fig:varyRes_noTransport} compares spectra for our $\beta = 10$ MHD simulations to simulations with $P_{CR} \sim P_{g}$ and purely advective CR transport (no streaming and $\kappa_{||} = 0$). For grid sizes of $256^{3}$ and $512^{3}$, we see in both cases that, in the inertial range up until $k \sim 20$, there is no appreciable damping due to the presence of CRs, confirming again that CR transport is the cause for clear and obvious damping seen in Figures \ref{fig:CR_Turb_Modify}, \ref{fig:KE_spectra_streaming}, \ref{fig:KE_spectra_varyKappa} beginning at small k. 

\begin{figure*}
\includegraphics[width=0.98\textwidth]{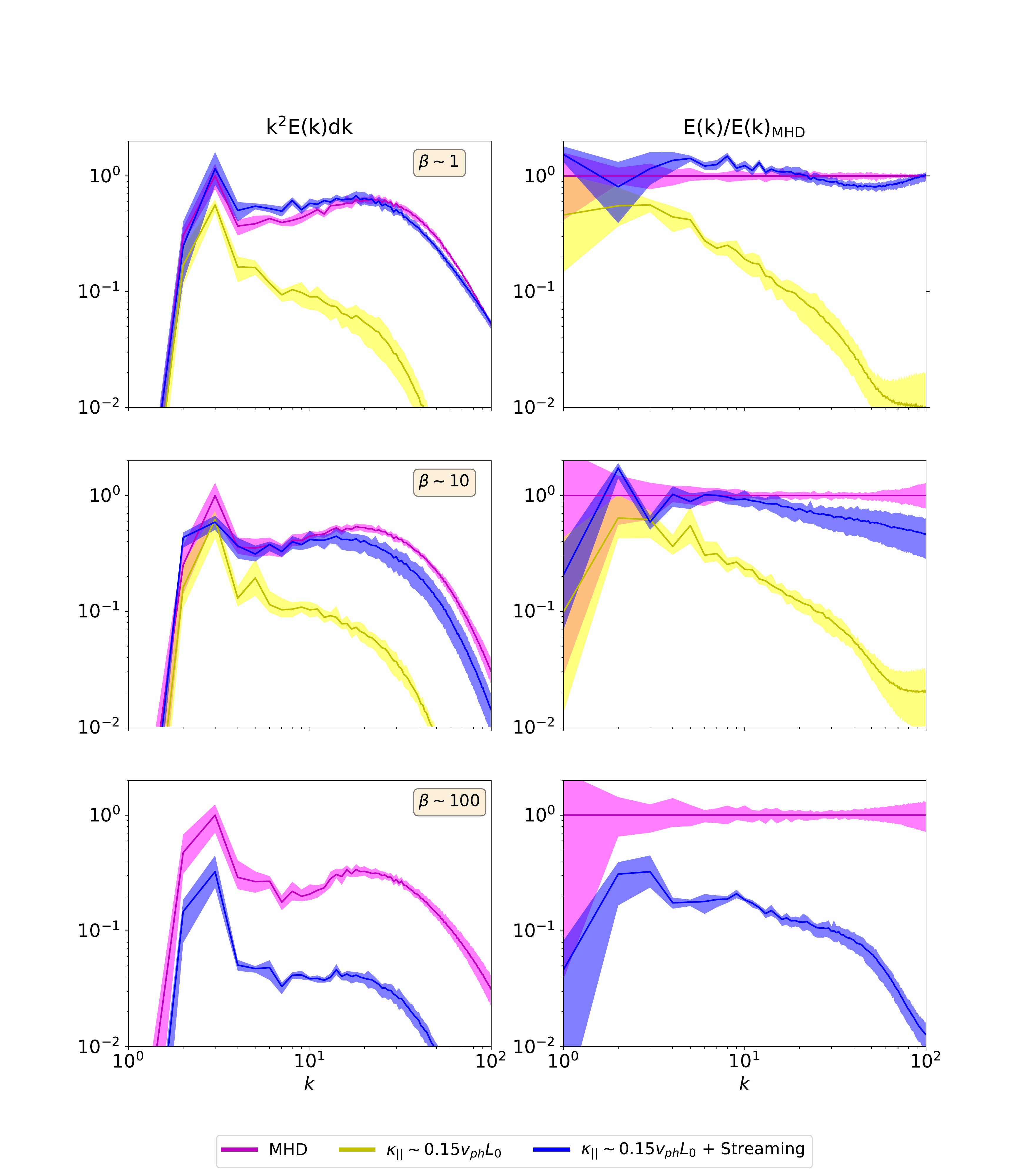}
\caption{Kinetic energy spectra for a partial simulation suite run on a  $512^{3}$ grid instead of a $256^{3}$ grid (compare to Figure \ref{fig:KE_spectra_streaming} in \S \ref{sec:streaming}). Computer resource limits prohibit us from running the streaming only ($\kappa \sim 0$ + streaming) simulations of \S \ref{sec:streaming} on a $512^{3}$ domain, but all other spectra look qualitatively similar to those on a $256^{3}$ grid; namely, diffusion only transport shows clear differences in spectral slope at both $\beta = 1$ and 10. Streaming simulations instead appear to uniformly decrease kinetic energy at all scales as $\beta$ increases. Note there is some resolution dependence for $\beta = 10$, with $512^{3}$ showing less damping compared to the $256^{3}$, but the difference is mild, especially in comparison to the diffusion-only simulations.}
\label{fig:streaming_spectra_res512}
\end{figure*}

Figure \ref{fig:streaming_spectra_res512} shows kinetic energy spectra for $512^{3}$ simulations when transport is included. These simulations only comprise part of those on a $256^{3}$ grid (compare to Figure \ref{fig:KE_spectra_streaming} in \S \ref{sec:streaming}) because computer resource limits prohibit us from running the streaming only ($\kappa \sim 0$ + streaming) simulations. In any case, the streaming only simulations and the streaming + diffusion simulations are both streaming dominated in this sub-Alfv\'{e}nic regime, so we expect their spectra to look very similar, as we saw in Figure \ref{fig:KE_spectra_streaming}.  

The MHD and diffusion only spectra look qualitatively similar to those on a $256^{3}$ grid. Most importantly, diffusive transport leads to significant damping compared to the MHD case at all $\beta$ tested ($\beta = $1 and 10). As in \S \ref{sec:streaming}, streaming transport instead appears to uniformly decrease kinetic energy at all scales, and this is $\beta$ dependent with $\beta \sim 1$ showing almost no difference between MHD and CR cases. There is some resolution dependence for $\beta = 10$, with $512^{3}$ showing less damping compared to the $256^{3}$ run, but the difference is mild, especially compared to the heavily damped diffusion-only simulations.

\bibliographystyle{apj}
\bibliography{bibliography, peng_citations}

\end{document}